\documentclass[b5paperpaper,12pt]{mn2e}
\pdfpagewidth=\paperwidth
\pdfpageheight=\paperheight
\pdfoutput=1
\usepackage[usenames,dvipsnames,svgnames,table]{xcolor}
\usepackage{graphicx, subfigure}  
\usepackage[pass]{geometry}
\usepackage{times}
\usepackage{float} 
\usepackage{aastex_hack}
\usepackage[export]{adjustbox} 
\usepackage[authoryear]{natbib}  
\usepackage{amsmath}
\usepackage{amssymb}
\usepackage{appendix}
\usepackage{setspace}
\usepackage{layout} 
\usepackage{wrapfig}
\usepackage{tabularx}
\usepackage{longtable}
\usepackage{bm}
\usepackage{lscape}
\usepackage{rotating}
\usepackage{multirow}
\usepackage{footnote}
\usepackage{array}

\begin{document}

\title{Detection of a Coherent Magnetic Field in the Magellanic Bridge through Faraday Rotation}
\author[J. F. Kaczmarek et al.]
{J. F.~Kaczmarek,$^1$\thanks{email: jane.kaczmarek@sydney.edu.au}
C. R.~Purcell,$^{1,2}$ B. M.~Gaensler,$^{1,3}$ N. M.~McClure-Griffiths,$^4$ 
\newauthor J.~Stevens$^5$\\
 	$^1$Sydney Institute for Astronomy, School of Physics, The University of Sydney, NSW 2006 Australia \\
 	$^2$Research Centre for Astronomy, Astrophysics, and Astrophotonics, Macquarie University, NSW 2109, Australia \\
 	$^3$Dunlap Institute, University of Toronto, 50 St. George Street, Toronto, ON M5S 3H4, Canada \\
 	$^4$Research School of Astronomy and Astrophysics, Australian National University, Canberra, ACT 2611, Australia \\
 	$^5$CSIRO Astronomy \& Space Science, 1828 Yarrie Lake Road, Narrabri, NSW 2390 Australia}
\maketitle

\begin{abstract} 
We present an investigation into the magnetism of the Magellanic Bridge, carried out through the observation of Faraday rotation towards 167 polarized extragalactic radio sources spanning the continuous frequency range of $1.3 - 3.1$\,GHz with the Australia Telescope Compact Array. Comparing measured Faraday depth values of sources `on' and `off' the Bridge, we find that the two populations are implicitly different. Assuming that this difference in populations is due to a coherent field in the Magellanic Bridge, the observed Faraday depths indicate a median line-of-sight coherent magnetic-field strength of $B_{\parallel}\,\simeq\,0.3\,\mu$G directed uniformly away from us. Motivated by the varying magnitude of Faraday depths of sources on the Bridge, we speculate that the coherent field observed in the Bridge is a consequence of the coherent magnetic fields from the Large and Small Magellanic Clouds being pulled into the tidal feature. This is the first observation of a coherent magnetic field spanning the entirety of the Magellanic Bridge and we argue that this is a direct probe of a `pan-Magellanic' field.

\end{abstract} 
\begin{keywords}
galaxies: Magellanic Clouds, galaxies: magnetic fields
\end{keywords}


\section{Introduction}

The Large Magellanic Cloud (LMC) and Small Magellanic Cloud (SMC) are a highly-studied galaxy pair. Due to their close proximity to the Milky Way (MW), the Magellanic Clouds allow astronomers to study galaxy interactions and evolution in unprecedented detail. The on-going interaction between the galaxy pair, and possibly the MW, have led to the creation of the Magellanic Bridge (MB), the Magellanic Stream, and the Leading Arm (see \citealt{Besla2010} and \citealt{D'OnghiaFox2016} for a complete review). Each of these tidal features can be identified through the presence of large amounts of H\textsc{i} gas. Most prominent of these features is perhaps the MB (\citealt{MB1963}) -- a contiguous, gaseous tidal feature that spans the region between the LMC and SMC. We assume that the MB is located at a distance of 55\,kpc, the mean distance to the LMC and SMC \citep{Walker1999}. We also assume that the bulk of the H\textsc{i} emission in the MB has a radial velocity in the range $+100\,$km\,s$^{-1}\leq\,v_{\text{LSR}}\leq\,+300$\,km\,s$^{-1}$ \citep{Putman2003, Muller2003}. The tidal remnant is thought to have formed $\sim$200\,Myr ago when the LMC and SMC were at their closest approach to one another \citep{GardinerNoguchi1996, Besla2012}.

Tidal tails, streams and bridges play an important role in the evolution of the parent galaxies as well as the host environment, as they serve as a siphon for galactic material to be dispensed into the diffuse intergalactic medium. It can be posited that a pre-existing magnetic field could follow the movement of neutral gas into the intergalactic medium. The stretching and compressing of tidally stripped gas may then serve as a mechanism for the amplification of any existing magnetic fields \citep{Kotarba2010}. Thus, the stripping of tidal debris may be partially responsible for the distribution of magnetic fields over large volumes. What remains unclear is the importance and role of magnetic fields within tidal features.

The association between tidal remnants and magnetic fields has been studied for nearly two decades. Classically, the radio continuum tidal bridge connecting the `Taffy' galaxies \citep{Condon1993} was estimated as having a similar magnetic-field strength to the pre-collision galaxies and the field lines appeared to be stretching across the space between the galaxy pair. More recently, tidal dwarfs within the Leo Triplet and Stephan's Quintet have been shown to possess coherent magnetic fields and have total magnetic-field strengths of $B_T\,=\,3.3\,\pm\,0.5\,\mu$G and $B_T\,=\,6.5\,\pm\,1.9\,\mu$G, respectively (\citealt{Nikiel2013a}a, \citealt{Nikiel2013b}b).

Decades of research using optical polarized starlight has shown that polarization vectors in the plane of the sky trace out a path from the SMC along the western Bridge oriented in the direction of the LMC \citep{MathewsonFord1970b, MathewsonFord1970a, Schmidt1970, Schmidt1976, Magalhaes1990, Wayte1990, LoboGomes2015}. Due to the limited number of stars with which one can carry out optical polarimetry studies, all previous claims of the existence of a coherent magnetic field spanning the entire Magellanic System have had to be speculative due to the lack of information stemming from the diffuse MB.

Studies of Faraday rotation of background polarized radio sources towards the LMC have determined that the galaxy has a coherent magnetic field of strength $\sim{1}\,\mu$G \citep{Gaensler2005}. \citet{Mao2008} observed the SMC using both Faraday rotation measures and polarized starlight. Through careful consideration of the Galactic foreground they constructed 3D models for the magnetic field and showed that the orientation of the field has a possible alignment with the MB.

A similar investigation into Faraday rotation towards extragalactic polarized sightlines has shown that a high-velocity cloud (HVC) in the Leading Arm hosts a coherent magnetic field \citep{McClureGriffiths-LA-2010}. In such an instance, a magnetic field could work to prolong the structural lifetime of the HVC as it is accreted onto the MW disk. While the exact origin of the magnetic field in this HVC remains unclear, it is plausible that the HVC fragmented from a magnetized Leading Arm. Therefore, the observed magnetic field in the HVC would be a consequence of the initial seed field followed by compression and amplification due to the MW halo.

Although magnetic fields have been found in the SMC, LMC, and some HVCs, none of the previous investigations of magnetism in the Magellanic System have directly confirmed the existence of the {\em Pan-Magellanic Field} -- a coherent magnetic field connecting the two Magellanic Clouds. 

\subsection{Faraday Rotation}
Complex linear polarization is an observable quantity and can be defined as
\begin{equation}\label{eq:compPol}
\mathcal{P}~=~Q + iU = p_0~e^{2i\Psi},
\end{equation}

\noindent where $Q,$ and $U$ are the observed linearly polarized Stokes parameters, $p_{0}$ is the polarization fraction intrinsic to the source and $\Psi$ is the observed polarization angle, also defined as:
 \begin{equation}\label{eq:Psi}
 \Psi = \frac{1}{2}~arctan~\frac{U}{Q}.
 \end{equation}
 
\noindent The polarization angle is rotated from its intrinsic value ($\Psi_{0}$) any time the emission passes through a magneto-ionic material. This effect is known as Faraday rotation. The total observed Faraday rotation, defined $\Delta\Psi/\Delta\lambda^2$, is known as the rotation measure (RM).

When the rotating material is located along the line-of-sight, Faraday rotation can serve as a powerful tool to analyse magnetism. In the simple case of a thermal plasma threaded by a single magnetic field, the intrinsic polarization angle is rotated by $\Delta\Psi~=~\text{RM}\lambda^{2}$ radians. However, recent studies have shown that the RM may offer an incomplete, or misleading diagnostic of the actual polarization properties along the line-of-sight \citep{O'Sullivan+2012, Anderson2016} and that many sources cannot be described by a single RM. It is therefore more robust to discuss the polarized signal in terms of its Faraday Depth ($\phi$), as first derived by \citet{Burn1966}. The Faraday depth encodes the electron density ($n_{e}$, in cm$^{-3}$) and magnetic-field strength along the line-of-sight ($B_{\parallel}$, in $\mu$G) according to
 \begin{equation}
\phi(L)~=~0.812\int^0_L~n_{e}B_{\parallel}dl,
\label{eq:RM}
 \end{equation}
\noindent where $L$ is the distance through the magneto-ionic material in parsecs. The sign of the Faraday depth is indicative of the orientation of the magnetic field with a positive $\phi$ signifying the field to be oriented towards the observer and a negative $\phi$ implying a field that is pointing away. 

The measured $\phi_{\rm{obs}}$ for a extragalactic source behind the MB is a summation of the various Faraday depth components along the line-of-sight and can be broken down into its constituent parts as follows:
\begin{equation}
	\phi_{\rm{obs}}\,=\,\phi_{\rm{intrinsic}}\,+\,\phi_{\rm{IGM}}\,+\,\phi_{\rm{MB}}\,+\,\phi_{\rm{MW}},
	\label{eq:phi_los}
\end{equation}
where $\phi_{\rm{intrinsic}}$ is the Faraday depth that is associated with the polarized emitting source, $\phi_{\rm{IGM}}$ is any rotation due to the intergalactic medium, $\phi_{\rm{MB}}$ is our targeted Faraday depth due to the posited MB magnetic field and $\phi_{\rm{MW}}$ is the Faraday rotation due to the foreground MW. Although $\phi_{\rm{intrinsic}}, \phi_{\rm{IGM}}$ and $\phi_{\rm{MW}}$ are present along all sightlines, $\phi_{\rm{MW}}$ is likely to dominate the observed signal. This assumption appears to have been well justified in \citet{TaylorMap}, whereby mapping the rotation measures of extragalactic polarized sources from the NRAO VLA Sky Survey (NVSS) revealed local structures in the Galaxy. Therefore, by observing polarized sources with sightlines that do not intersect the MB, we will be able to correct for the Galactic foreground, leaving the residual $\phi$ to represent the intrinsic properties of the background source and the MB contribution. The intrinsic polarized properties of each polarized source are random and considered to have a negligible effect on the overall statistics for a large sample.

\begin{figure}
\includegraphics[width=1\linewidth]{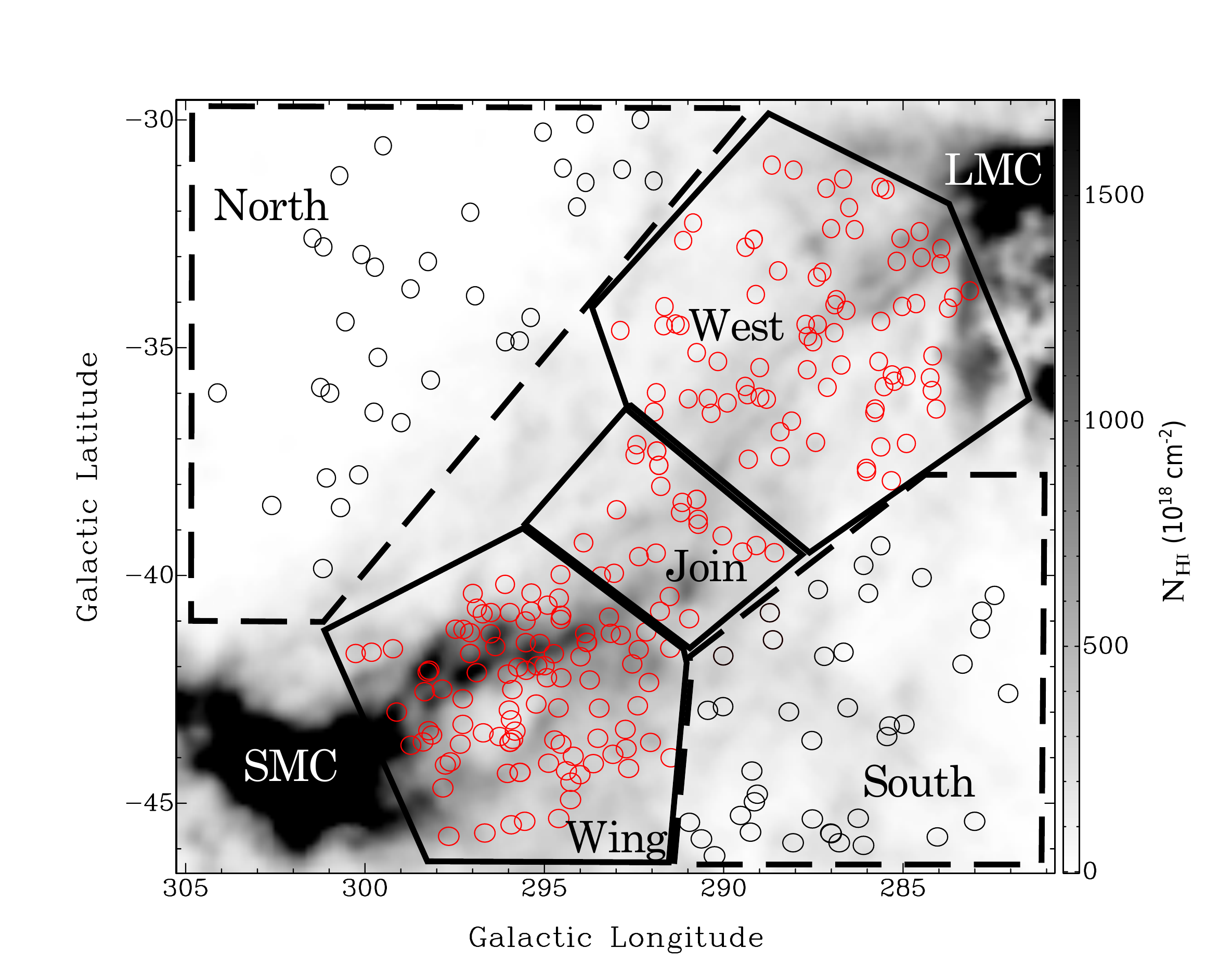}
\caption{Neutral hydrogen column density for the velocity range of $+100\,\leq\,v_{\rm{LSR}}\leq+300$\,km\,s$^{-1}$ of the MB region from the GASS survey (\citealt{GASS}, \citealt{GASSIII}), over-plotted with the positions of observed radio sources. Each pointing is associated with a region name denoted by the text in the enclosed areas. Red circles (pointings enclosed by a solid line) are sources where the MB is considered to intersect the background source's line-of-sight, whereas sources marked by black circles (pointings enclosed by a dashed line) are considered as having lines-of-sight that are not contaminated by the MB. These latter sources were observed in order to subtract the Faraday rotation contribution from foregrounds and backgrounds. }
\label{fig:allPts}
\end{figure}

\begin{table*}
\caption{Summary of the observations. Column 1 gives the array configuration; Column 2 gives the regions targeted (as defined in $\S$2) ; Column 3 lists the length of the observing run and Column 4 gives an approximation for the total integration time per source. Column 5 gives the UT date of the commencement of the observations.} 
\footnotesize 
\centering 
\begin{tabular}{ l l l l l }
Array Config. & Obs. Targets 						& Obs. Length	& Time On-Source 	& Obs. Date 	 \\ 
							& 												& (hrs) 					& (min) 						& 						 \\ 
\hline 
6C 	 					&  Wing, West 						&	12 						&	2.5								&	2015 Mar 14 \\ 
6A						&	 Wing, West							&	15						& 1.5								&	2015 Apr 30 \\
6A						&	 Join, North, South			&	15						& 	3								&	2015 Apr 30 \\
1.5B 					&  Wing (subset)					&  3 						& 	5								& 2016 Jun 11 \\ 
\hline
 \end{tabular}
 \end{table*}

If there exists a coherent magnetic field threading the MB, observations of linearly polarized background radio sources may hold the key to its discovery. In this work, we use detailed measurements of the Faraday depth of background, extragalactic polarized sources to investigate the existence of a coherent magnetic field spanning the MB. We describe our source selection process and observations in Section\,\ref{sec:obs}, followed by data reduction and processing in Section\,\ref{sec:redux}. We present our results in Section\,\ref{sec:results}, which include the fitting and subtraction of the MW foreground. Section\,\ref{sec:B_strengths} motivates different distributions of ionized gas and the subsequently derived magnetic-field strengths. In Section\,\ref{sec:discuss} we discuss the possible origins and implications of the pan-Magellanic Field. A summary is presented in Section\,\ref{sec:conclude}.

\section{Observations \& Data}
\label{sec:obs}

\subsection{Source Selection}
For this investigation, we observed a subset of polarized sources that were originally identified through the reduction and re-processing of archival continuum data of the western MB (see \citet{Muller2003} for a summary of observations). In the literature, this region has been referred to usually as either the `Wing' or `Tail' \citep{Bruns2005, Lehner2008}, and we make reference to this region as the `Wing,' exclusively (See Figure\,\ref{fig:allPts} for location). The H\textsc{i} observations of the `Wing' had simultaneously observed the continuum emission associated with this region. The source-finding algorithm \textit{Aegean} \citep{aegean} was used to identify polarized sources in the final, deconvolved continuum images. From this original sample, we targeted 101 polarized sources for follow-up observations. 

An additional 180 radio sources were targeted in order to extend the investigation across the entirety of the MB and surrounding area. Motivated by the changing morphology and kinematics of the Bridge, we separate these additional sources into regions `West', 'Join', `North', and `South'. These additional radio sources were selected from the Sydney University Molonglo Sky Survey (SUMSS, \citealt{SUMSS}) as having a Stokes $I$ flux $\geq\,100\,$mJy at 843\,MHz for the region labelled `West' and $\geq\,150$\,mJy for regions `Join,' `North' and `South'. Figure \ref{fig:allPts} gives a summary of the pointing regions observed overlaid on a map of neutral Hydrogen (H\textsc{i}) of the region from the Galactic All Sky Survey (GASS; \citealt{GASS, GASSIII}). 

\subsection{Observations}
Observations of the 281 radio sources were taken over 3 days with the Australia Telescope Compact Array under project C3043. Taking advantage of the instantaneous broad bandwidths of the Compact Array Broadband Backend (CABB, \citealt{CABB}), the observations spanned the continuous frequency range of 1100\,--\,3100\,MHz. Each pointing was observed as a series of snapshots in order to improve $uv$-coverage. Phase calibrators were observed at least every 40 minutes. The bandpass and flux calibrator PKS\,B1934-638 was observed on 14 March 2015 and 30 April 2015 and PKS\,B0823-500 was observed as the bandpass calibrator on 11 June 2016. polarization leakage calibrations were carried out using the aforementioned primary calibrators. On average, each pointing was observed for a total of 3 minutes. Due to the nature of the source selection associated with the `Wing' and the possibility that sources could be weak in total intensity, the initial 3 minutes of observation was sometimes not enough to reach a sufficient signal-to-noise. Additional observations were made as a single hour-angle $uv$-cut on 11 June 2016 in order to improve our sensitivity limits for points that were not bright enough in polarization nor total intensity to be confidently detected with our initial observations. A summary of the observations is listed in Table 1 and the representative $uv$-coverage for any source in each region is shown in Figure 2.

\begin{figure*}
\centering
\subfigure[Wing and West regions]{\label{fig:uv-wing}\includegraphics[width=0.48\textwidth]{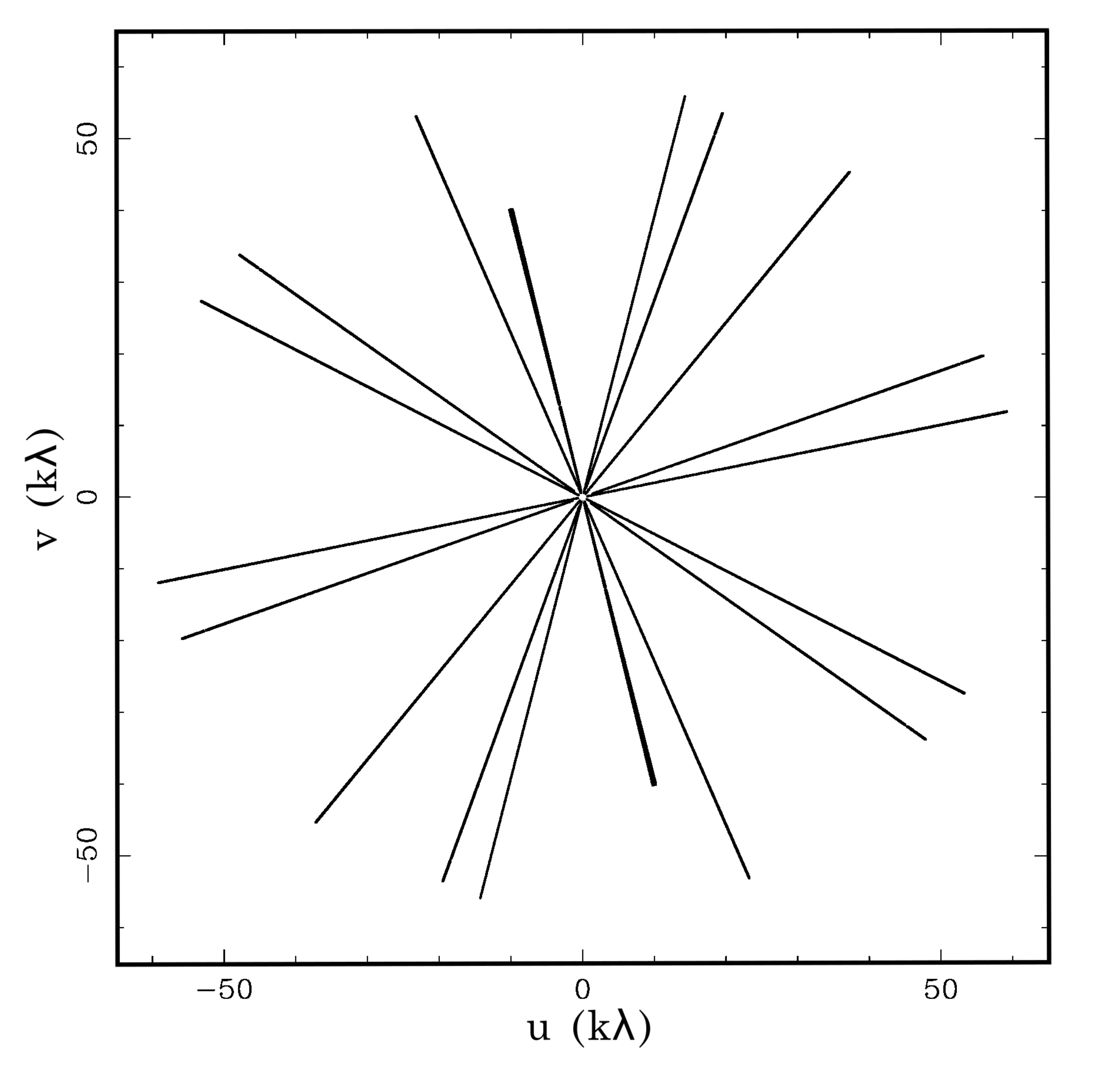}}
\hfill
\subfigure[Join, North and South regions]{\label{fig:uv-join}\includegraphics[width=0.48\textwidth]{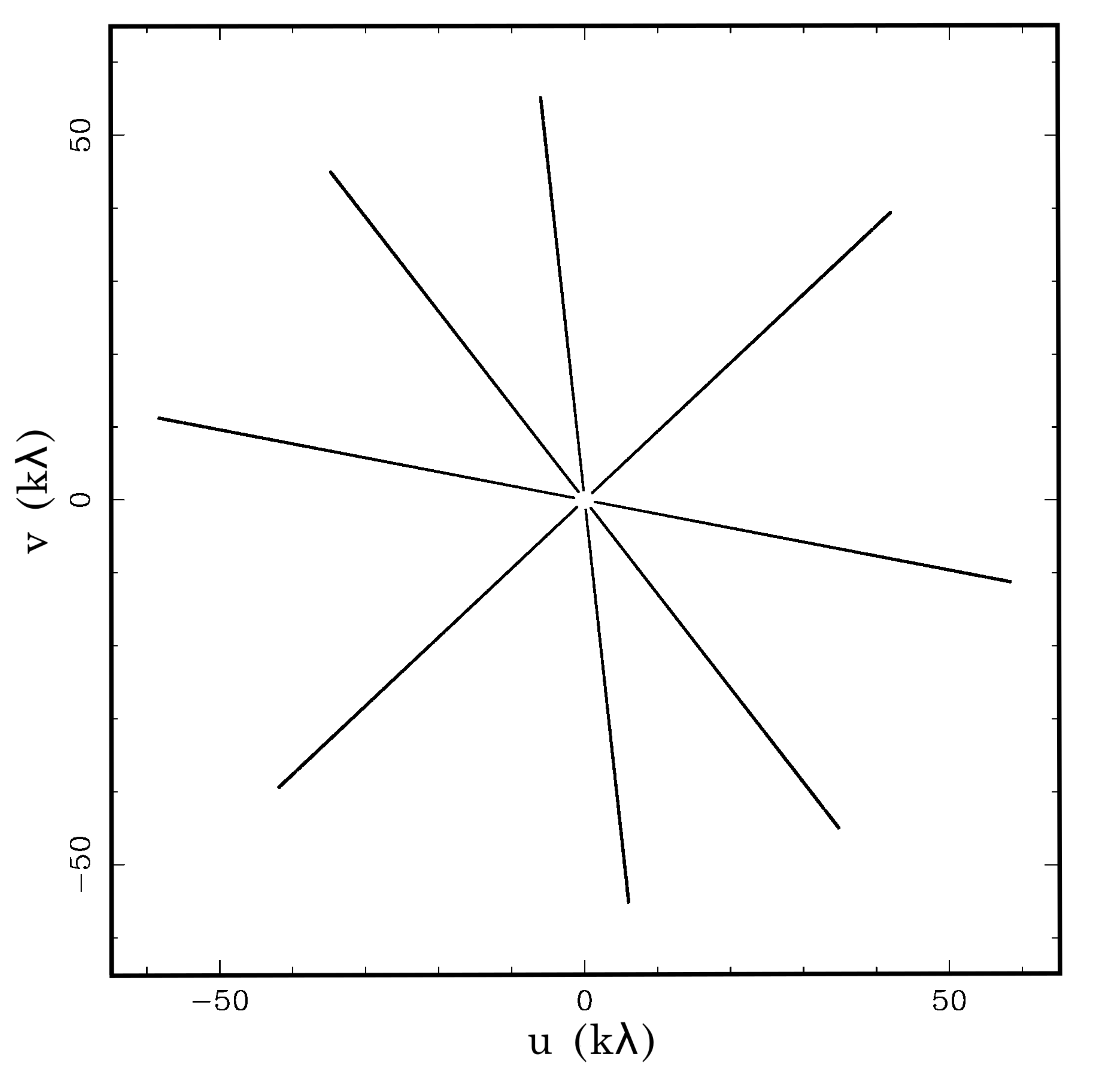}}
\caption{Typical $uv$-coverage of a single radio source associated with (a) the `Wing' and `West' and (b) `Join', `North', `South'. }
\end{figure*}
 
 \section{Data Reduction and Extraction}
 \label{sec:redux}
Observations were calibrated and imaged in the {\sc miriad} software package \citep{miriad} using standard routines. Flagging of the data was done largely with the automated task \textsc{pgflag}, with minor manual flagging being carried out with tasks \textsc{blflag} and \textsc{uvflag}. Naturally-weighted Stokes $I$, $Q$, $U$ and $V$ maps were made using the entire 2\,GHz bandwidth. Deconvolution of the multi-frequency dataset was performed on the dirty maps with the task \textsc{mfclean}. Cleaning thresholds were set to be 3 times the rms Stokes $V$ levels (3$\sigma_V$) for Stokes $Q$ and $U$, and 5$\sigma_V$ for Stokes I. Images were convolved to a common resolution of 8\,arcseconds, which corresponds to a linear scale of 2\,pc at the assumed distance to the MB of 55\,kpc. 

From the broadband 2\,GHz images, images of linearly polarized intensity ($\mathcal{P}$) were made with the task \textsc{math}. The total polarized flux of a target was extracted from an aperture 8\,arcseconds in diameter centred on the peak polarization pixel with noise estimates ($\sigma_{\mathcal{P}}$) measured as the rms residuals from a source-extracted image. A target was considered `polarized' if the integrated polarized flux was greater than 8$\sigma_{\mathcal{P}}$. This method of imaging will lead to bandwidth depolarization for sources with absolute Faraday depths greater than $\sim90\,$rad\,m$^{-2}$; however, we consider the number of sources rejected due to high Faraday rotation to be negligible and has no impact on our final science goals.

 \begin{table}
 \centering
 \begin{tabular}{l c c c}
 Region			&		Observed		&		Polarized		&		Accepted fraction 	\\
 						&								&								&			(\%)					\\
 \hline\hline
 Wing				&			101				&			69				&			68						\\
 West 			&			83				&			40				&			48						\\
 Join				&			23				&			15				&			65						\\
 North			&			34				&			22				&			65						\\
 South			&			40				&			21				&			53						\\
 \hline
 						&			281 			&			167				&			59						\\
 \end{tabular}
\caption{Summary of total number of points observed per region and total number of polarized sources accepted. In order to be accepted, a source must be detected to at least 8$\sigma$ in the full bandwidth polarized intensity image. The `Wing' region returns a higher fraction of polarized sources due to our previous knowledge of the polarization in this region.} 
\label{table:acceptedFrac}
\end{table}

\begin{figure*}
\centering
\subfigure[Total intensity for source Join\_08]{\includegraphics[width=0.48\textwidth]{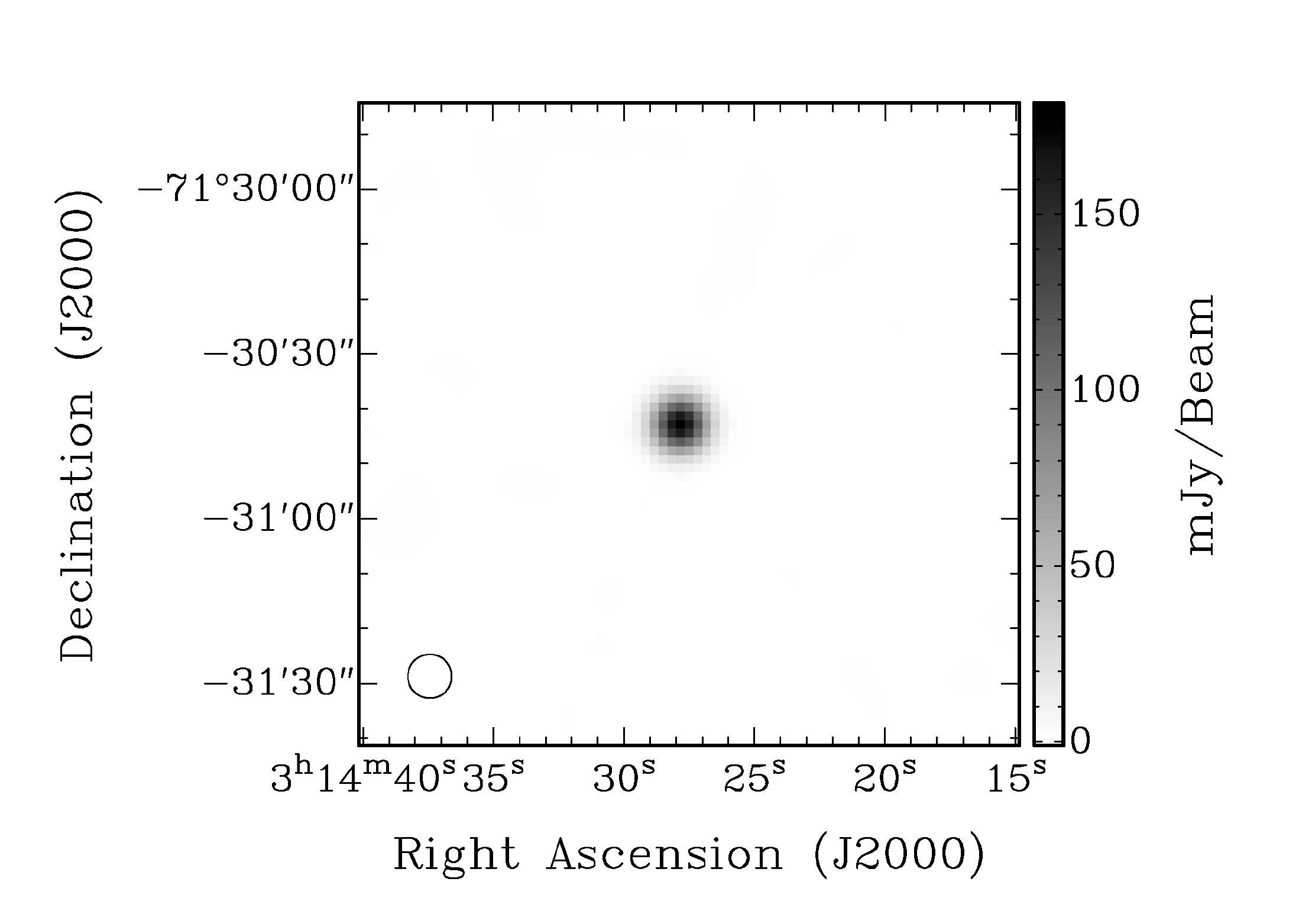}}
\hfill
\subfigure[Polarized intensity for source Join\_08]{\includegraphics[width=0.44\textwidth]{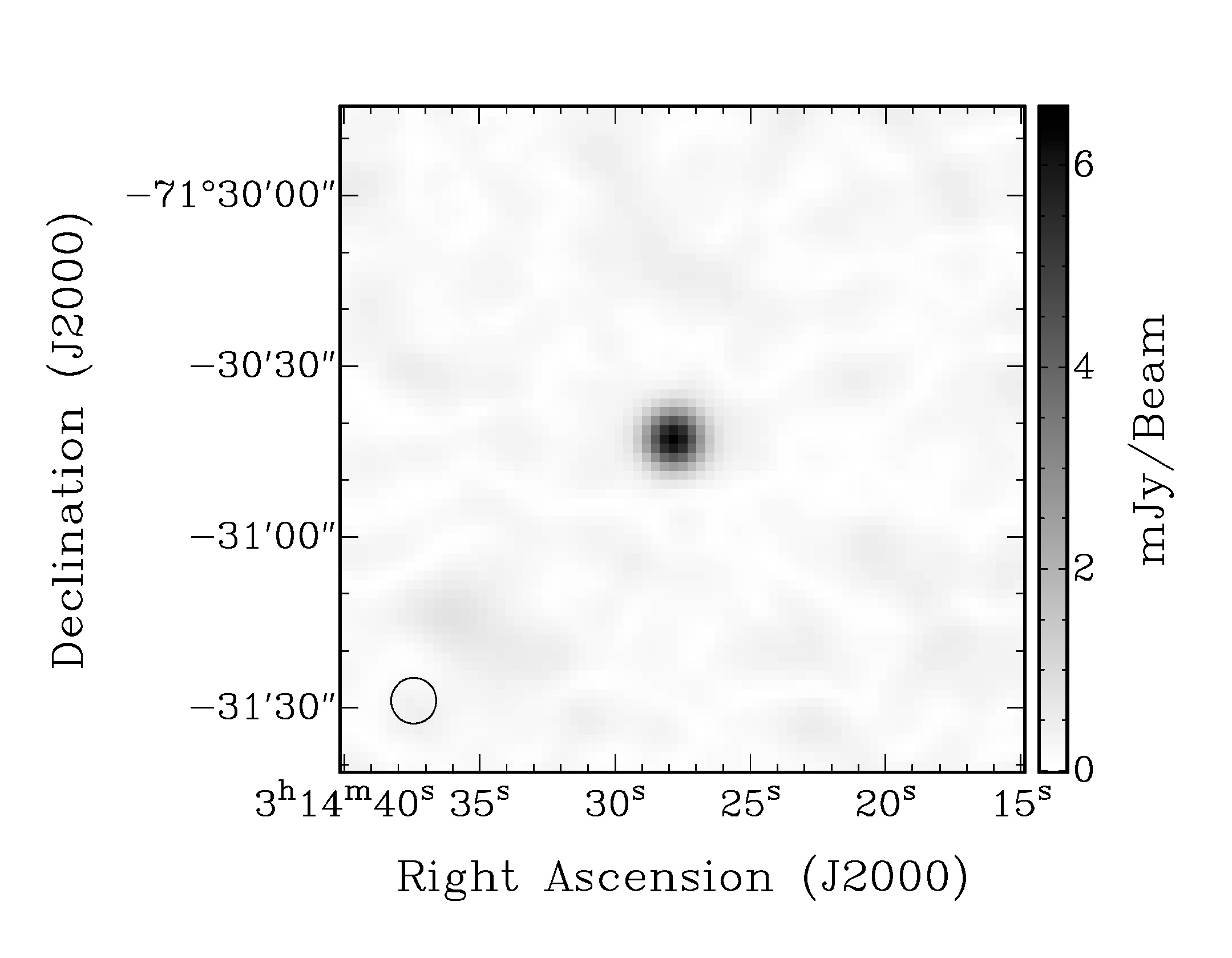}}
\subfigure[Total intensity for source West\_02]{\includegraphics[width=0.48\textwidth]{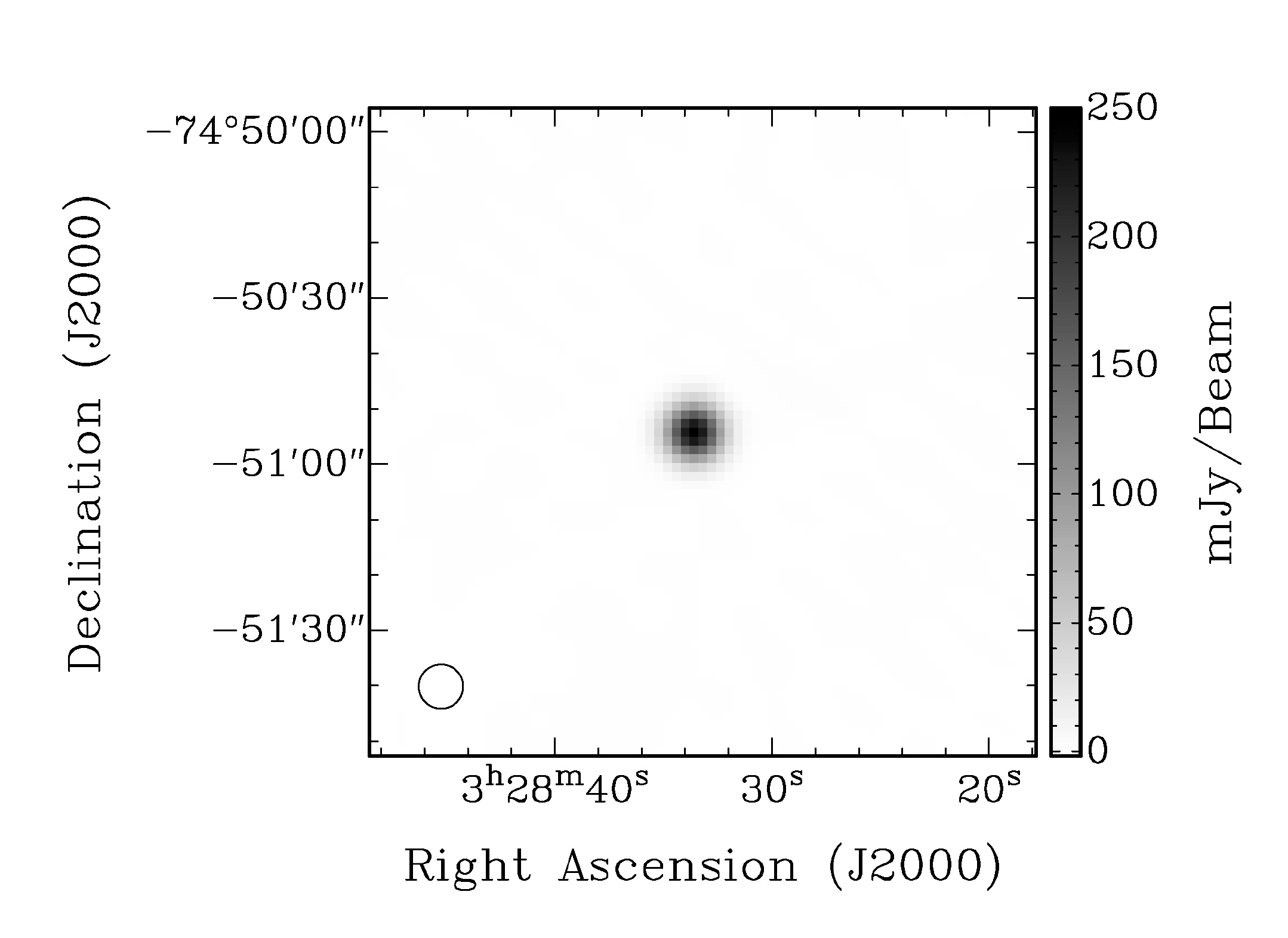}}
\hfill
\subfigure[ polarized intensity for source West\_02]{\includegraphics[width=.44\textwidth]{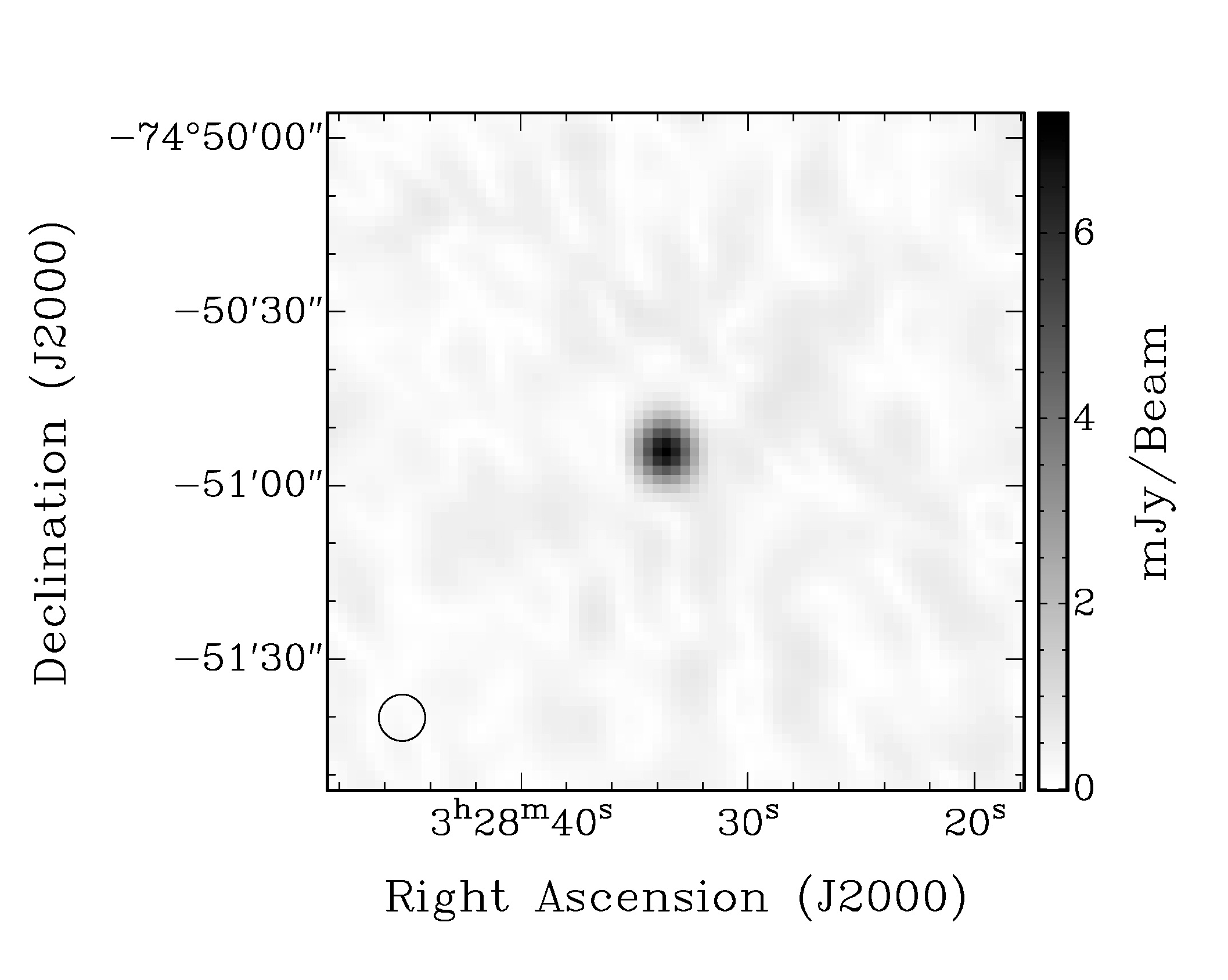}}
\caption{Example of two polarized sources detected in our survey: points `Join\_08' (a) and (b) and `West\_02' (c) and (d). Multi-frequency images for total intensity (Stokes\,$I$) are shown in (a) and (c) and polarized intensity ($\mathcal{P}$) in (b) and (d). Both sources have been imaged using the full bandwidth available and the restoring beam is shown in the bottom left of each image. } 
\label{fig:ptSources}
\end{figure*}

Imaging with narrow bandwidths decreases the signal-to-noise in addition to reducing the resolution in Faraday depth space, while broad bandwidths decrease the maximum observable scale in Faraday space, as well as the maximum observable Faraday depth. In order to minimise the bandwidth depolarization and maintain a desirable signal-to-noise ratio, Stokes $I$, $Q$, $U$ images were made every 64\,MHz - resulting in 27 channel maps spanning 1312 - 3060\,MHz. 

 As with the broadband $\mathcal{P}$ images, integrated fluxes were extracted from each map from an equivalent beam area centred on the pixel corresponding to the peak in $\mathcal{P}$. Error measurements were estimated as the rms-noise level from images created from the residual of the Stokes maps after the source aperture was blanked. With the exception of sources associated with the `Wing', all targets are expected to be bright in total intensity. A further 10$\sigma$ cut-off was imposed, and extracted spectra with fewer than 10 channels were discarded. 

The procedures described above result in 167 sources with spectra in $I$, $Q$ and $U$. Table\,\ref{table:acceptedFrac} has a summary of the fraction of sources accepted per region. The `Wing' region has an advantage in returning a higher number of polarized sources due to our previous knowledge of the polarized detections in the region. However, our data extraction method rejected multiple targets in the `Wing' region for falling below the sensitivity threshold. Figure\,\ref{fig:ptSources} gives two examples of total ((a) and (c)) and polarized intensity ((b) and (d)) detected from extragalactic radio sources. 

\begin{table*}
\centering
\caption{A subset of measured and calculated source parameters. Columns\,(1) and (2) give the source location in Galactic longitude and latitude, respectively. Column\,(3) lists the integrated total intensity ($I$) over the full 2\,GHz bandwidth with uncertainties. Integrated polarized flux ($\mathcal{P}$) with uncertainty is listed in Column\,(4). Columns (5 - 8) give the best-fit parameters returned from $qu$-fitting: namely, the intrinsic polarization fraction (Column (5)), the intrinsic polarization angle (Column (6)), the total Faraday depth along the line-of-sight (Column (7)) and the Faraday dispersion (Column (8)).}
\begin{tabular}{ c c | c c | c c c c }
(1)		&		(2)		&		(3)		&		(4)		&		(5)		&		(6)		&		(7)		&		(8)		\\
	$l$	&		$b$		&		$I$		&	$\mathcal{P}$	&	$p_0$	&	$\Psi_0$	&	$\phi_{\rm{obs}}$		&		$\sigma_{\phi}$		\\
 ($^{\circ}$)	&	($^{\circ}$)	&	(mJy)		&	(mJy)			&		(\%)	&	($^{\circ}$)	&	rad\,m$^{-2}$	&		rad\,m$^{-2}$	\\
\hline\hline
&&&&&&&\\[-2mm]
	291.778	& -40.785	&	$104.9\,\pm\,0.2$	&	$3.2\,\pm\,0.3$ 	&	$5.8^{+0.7}_{-0.6}$			&	$37^{+4}_{-4}$				&		$+6^{+3}_{-4}$				&		$21^{+2}_{-2}$				\\[1mm]
	288.589	&	-39.501	&	$258.4\,\pm\,0.07$	&	$3.9\,\pm\,0.3$		&	$1.66^{+0.09}_{-0.08}$	&	$82^{+3}_{-3}$				&		$-0.2^{+2}_{-2}$			&		$3^{+2}_{-2}$					\\[1mm]
	290.958	&	-45.418	&	$215\,\pm\,3$			&	$7.1\,\pm\,0.4$		&	$2.8^{+0.3}_{-0.3}$			&	$49^{+5}_{-5}$				&		$+13^{+4}_{-4}$			&		$23^{+2}_{-2}$				\\[1mm]
	285.625	&	-39.347	&	$123.8\,\pm\,0.2$	&	$1.6\,\pm\,0.2$		&	$6.38^{+0.07}_{-0.07}$	&	$143.9^{+0.5}_{-0.5}$	&		$+26.8^{+0.3}_{-0.3}$	&		$13.9^{+0.2}_{-0.2}$	\\[1mm]
  296.659	&	-45.653	&	$73.0\,\pm\,0.2$		&	$6.8\,\pm\,0.3$		&	$10.1^{+0.4}_{-0.4}$		&	$71^{+2}_{-2}$				&		$-13.1^{+0.9}_{-1}$		&		$10.7^{+0.8}_{-0.8}$	\\[1mm]
\hline
\end{tabular}
\label{table:pts_intrinsic}
\end{table*}

\begin{figure}
\centering
\subfigure[Fractional Stokes spectra]{\includegraphics[width = 1\linewidth]{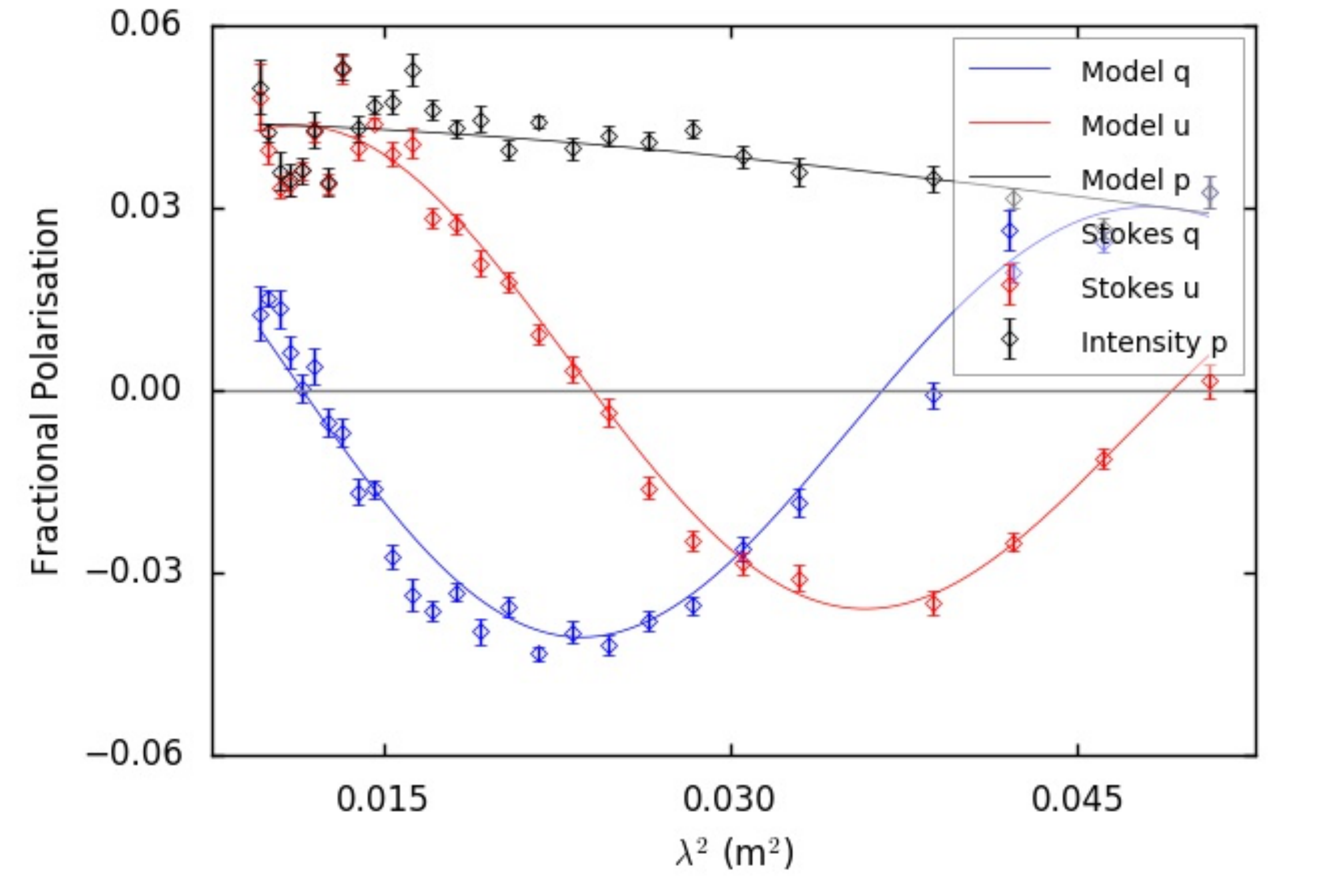} }
\centering
\subfigure[$\Psi$ vs. $\lambda^{2}$]{\includegraphics[width=.97\linewidth]{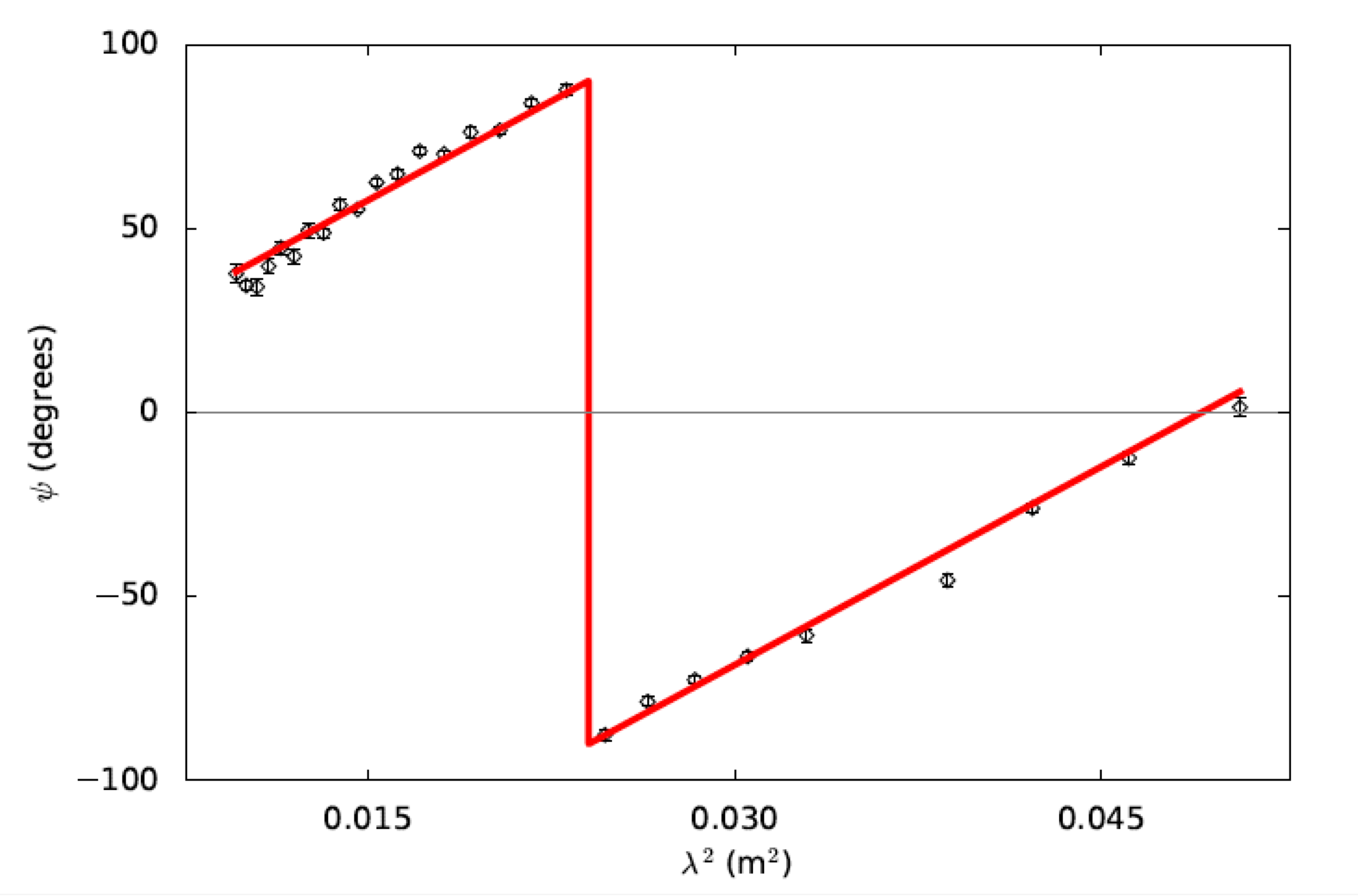}}	
\caption{ (a) Observed data and best-fit solution for $qu$-fitting to a point in the `West' region. Observed fractional Stokes $q$ and $u$ are shown as blue and red points, respectively, whereas the model solution is shown as blue and red lines. The observed and model polarized fraction is shown as black points and a black line for reference. (b) Corresponding fit to polarization angle ($\Psi$) versus $\lambda^{2}$ for the aforementioned solution from $qu$-fitting.}
\label{fig:quFit}
\end{figure}

\subsection{$qu$-fitting and $\phi$ determination}

We adopt the fractional notation such that $q\,=\,Q/I$ and $u\,=\,U/I$, where the observable polarized fraction can be expressed as
\begin{equation}
p~=~\sqrt{q^{2} + u^{2}}.
\end{equation}
In working with fractional Stokes parameters the wavelength dependent depolarization effects are decoupled from spectral index effects. 

To create our fractional polarized spectra, the $Q$ and $U$ spectra are divided by a model fit to the Stokes $I$ spectrum. This approach avoids creating non-Gaussian noise and the propagation of small-scale spectral errors that may be present in the Stokes $I$ spectrum. Using a bootstrap approach with 10,000 iterations, we fit a second-order polynomial to the Stokes $I$ spectrum of each polarized source and calculate the standard deviation of the resultant $q$ and $u$ values for each frequency channel. The total error is considered to be the standard deviation of the bootstrapped values of $q$ and $u$ added in quadrature to the measured noise from the cleaned Stokes $Q$ and $U$ maps. The bootstrap method is necessary to correctly propagate the uncertainty due to the fit and has the overall effect of increasing the magnitude of the errors from what can be measured from the Stokes maps. 

In order to extract the observed Faraday depth from our polarized signal, we must motivate a polarization model for the MB environment. External Faraday dispersion \citep{Burn1966} can be used as a proxy to measure fluctuations in the free-electron density or magnetic-field strength. This model has been used in numerous past studies of the polarization of galaxies, galaxy groups and clusters \citep{Laing+2008_structures, Gaensler2005}. Without an observed continuum-emission component of the MB, a single-component external Faraday dispersion model serves as an appropriate approximation to the polarization signal associated with the MB.

Polarization of this form displays a decreasing polarization fraction as a function of $\lambda^2$. This depolarization can be defined as $p/p_{0}$, where $p$ is the observed polarization. This effect is most evident towards long wavelengths. Due to the purely external dependence of external Faraday dispersion, and its dependence on the size of the observing beam, this depolarization model is often referred to as \textit{`beam depolarization'}. In this scenario, averaging the fluctuations across the entire beam area, the result is polarization of the form 
\begin{equation}
\mathcal{P}~=~p_0~e^{2i(\Psi_0+\phi_{\rm{obs}}\lambda^2)}~e^{-2\sigma^{2}_{\phi}\lambda^4},
\label{eq:EFD}
\end{equation}
\noindent where $\phi_{\rm{obs}}$ is the total observed Faraday-depth value (Equation\,\ref{eq:phi_los}) and $\sigma_{\phi}^2$ characterises the variance in Faraday depth on scales smaller than our beam. 

We calculate the best-fit $\phi_{\rm{obs}}$, $\sigma_\phi$ and $\Psi_0$ for each point source by fitting an external Faraday dispersion model (Equation\,\ref{eq:EFD}) simultaneously to the extracted $q(\lambda^{2}$) and $u(\lambda^{2}$) data. This technique is called $qu$-fitting and \citet{Sun2015} show it to be the best algorithm currently available for minimising scatter in derived polarization parameters. We take a Monte Carlo Markov chain (MCMC) approach to fitting our complex polarization parameters by employing the`emcee' \textit{Python} module \citep{emcee}. Compared to Levenburg-Marquardt fitting, MCMC better explores the parameter space, and returns numerically-determined uncertainties for the model parameters. The log-likelihood of the complex polarization model of the joint $q − u$ chi-squared ($\chi^{2}$) is minimised to find the best-fitting parameters. For each pointing, we initialise a set of 200 parallel samplers that individually and randomly explore the $n$-dimensional parameter space (where $n$ is the degrees of freedom). Each of these samplers -- called `walkers' -- iteratively calculate the likelihood of a given location in parameter space and in doing so map out a probability distribution for a set of parameters. 
 
We initialise the walkers to random values of the free parameters and run three 300 iteration `burn-in' phases where the samples settle on a parameter set of highest likelihood. The position history of the walkers is removed before initiating a 300-step exploration of the new parameter sub-space. The best fit model is calculated as the mean of the marginalised posterior distribution for each parameter. The parameter uncertainties are measured from the 1$\sigma$ deviation of the walkers above and below the resultant best-fit. 

Figure\,\ref{fig:quFit} gives an example solution from $qu$-fitting. The fractional Stokes spectra ($p$, $q$ and $u$) versus $\lambda^{2}$ is shown in the top panel (a). Observed values are shown as black, blue and red points for $p, q$ and $u$, respectively. The best-fit solution is shown to trace the observed data. The best fit solution to $\Psi$ versus $\lambda^{2}$ is given in the bottom panel (b). We attribute any deviation from the model to Faraday complexity of the source or a line-of-sight component that is not accounted for in the simple polarization model we assume (Equation\,\ref{eq:EFD}).

\section{Results}
\label{sec:results}

\begin{figure*}
\includegraphics[width = \textwidth,trim = {20 50 0 20}, clip]{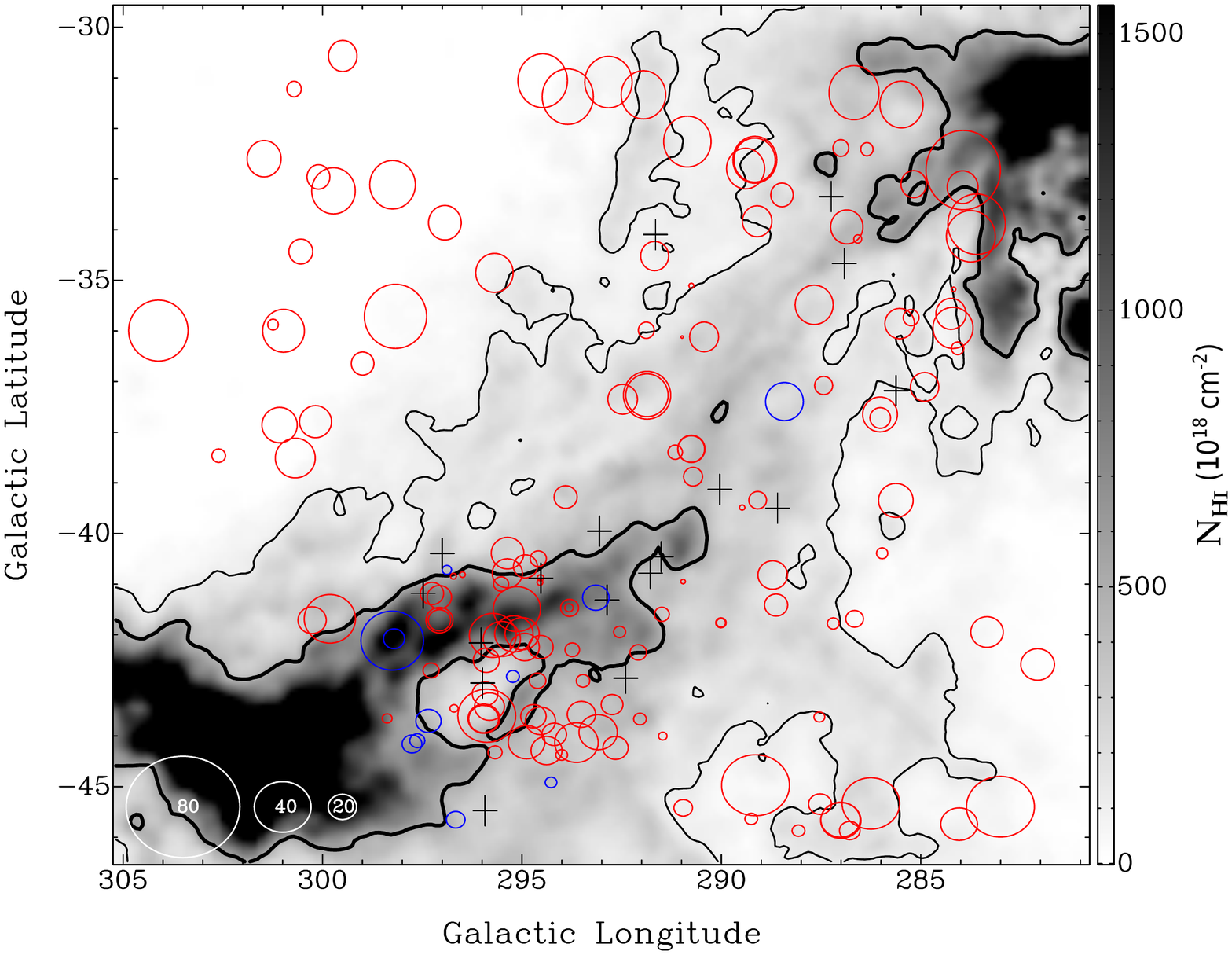}
\caption{$\phi_{\rm{obs}}$ values fit to an external Faraday dispersion model overlaid on a map of H\textsc{i} intensity from GASS \citep{GASSIII} in the velocity range of $+100\,\leq\,v_{\rm{LSR}}\,\leq\,+300$\,km\,s$^{-1}$. Black contours represent H\textsc{i} emissivity of 1.2 and 5.0$\times10^{20}$\,cm$^{-2}$. The size of each circle is representative of the magnitude of $\phi$, with scale-circles shown in the bottom left corner. Red circles represent a line-of-sight magnetic field pointing towards the observer (positive $\phi$), and blue circles show a field that is pointing away (negative $\phi$). Black crosses show $\phi$ values consistent with zero to 2$\times{d}\phi$.}
\label{fig:rawRMs}
\end{figure*}

In addition to fitting the observed Faraday depth ($\phi_{\rm{obs}}$), our fitting routine also returns best-fit values for all polarization parameters defined in Equation\,\ref{eq:EFD}, namely $p_0$, $\Psi_0$ and $\sigma_{\phi}$. A subsample of sources with the resultant best-fit parameters is given in Table\,\ref{table:pts_intrinsic}, with the full dataset available in Appendix \ref{sec:apndxB} .

Figure\,\ref{fig:rawRMs} shows the best-fit $\phi_{\rm{obs}}$ of every polarized radio source plotted over the H\textsc{i} emission of the region from GASS \citep{GASS, GASSIII}. Red circles indicate a positive $\phi_{\rm{obs}}$ and a field that is oriented towards the observer; blue circles, the opposite. Black crosses signify a $\phi_{\rm{obs}}$ that is consistent with zero to 2$\times{d}\phi$ where $d\phi$ is the returned uncertainty in Faraday depth from $qu$-fitting. 

We divide the observed polarized sources into two populations -- those where the MB intersects the sightline to the polarized source and those with sightlines that are unaffected by the MB. We define an `on-Bridge' region to be the area defined by a non-extinction corrected H$\alpha$ intensity of $I_{\text{H}\alpha}$\,=\,0.06\,R, shown as the lowest contour in Figure\,\ref{fig:corrRMs}. The H$\alpha$ dataset and subsequent analysis is discussed in more detail in Section\,$\S$\ref{sec:Ha_and_EM}. All sources associated with the `Wing,' `West' and `Join' regions meet this criterion. The `North' and `South' regions are considered to be `off-Bridge' and serve as a probe of the MW's Faraday depth structure in the region.

\begin{figure}
\includegraphics[width = 0.9\linewidth]{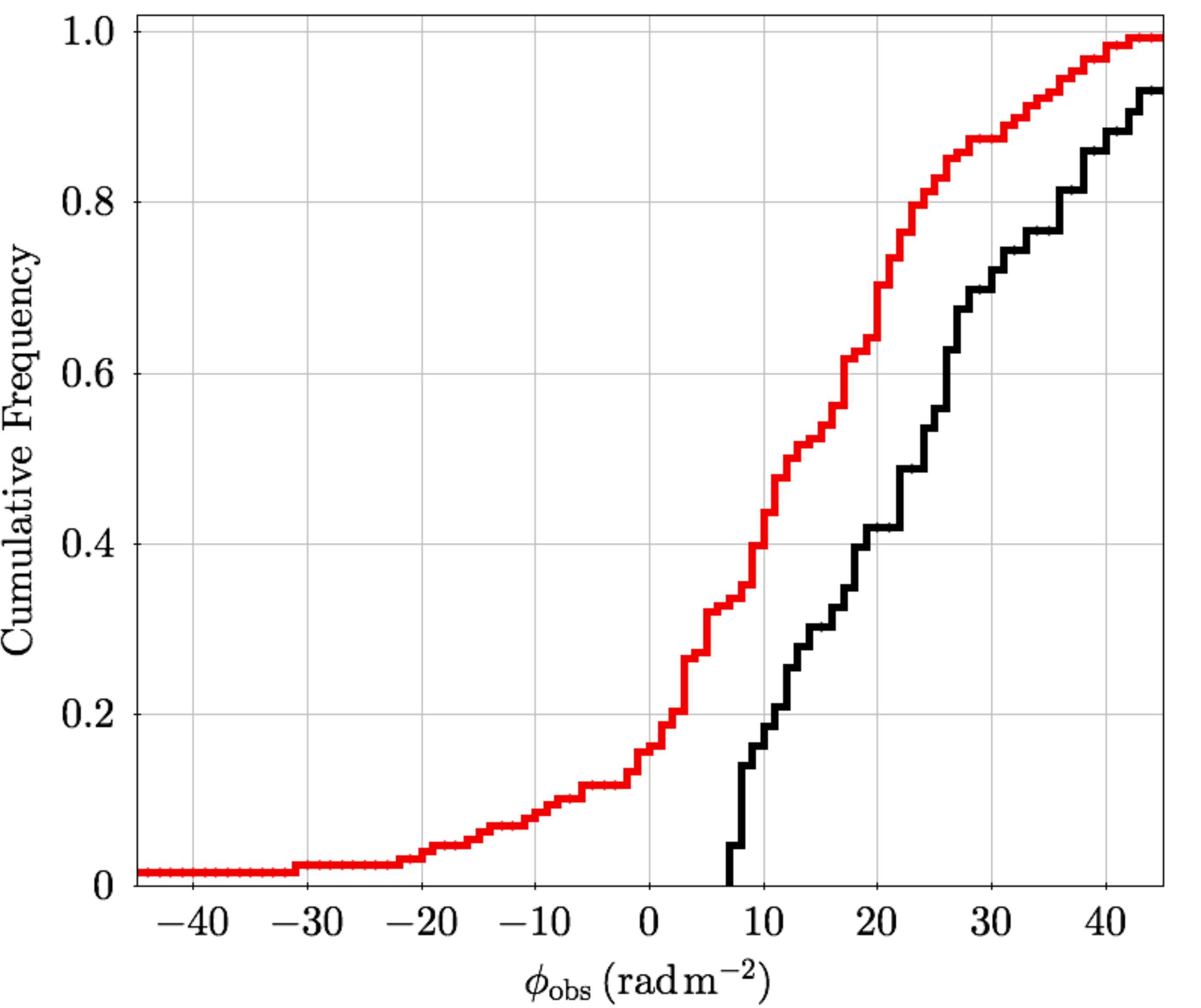}
\caption{Cumulative histogram of RM values for on-Bridge (red) and off-Bridge (black) sources. The figure is truncated at $\phi\,=\,\pm\,45\,$rad\,m$^{-2}$ for clarity.}
\label{fig:cumHist}
\end{figure}

Of all the $\phi_{\rm{obs}}$-values in the imaged region, $84\%$ are positive (red), and all of the negative (blue) and null (cross) Faraday depths are associated with the on-Bridge region (Figure\,\ref{fig:rawRMs}). Figure\,\ref{fig:cumHist} shows the $\phi_{\rm{obs}}$ population of all on- and off-Bridge sources as a cumulative histogram and highlights the clear discrepancy in Faraday depths for each population. We test the statistical likelihood that the Faraday depths associated with points on and off the Bridge come from a single population by performing a K-sample Anderson-Darling test on the best-fit $\phi_{\rm{obs}}$-values for all sources that have been detected to 8$\sigma_{\mathcal{P}}$ or higher in polarized intensity. The returned normalised test statistic allows us to reject the null hypothesis with a 99.992\% confidence level. The difference in Faraday depths between the populations of $\phi_{\rm{obs}}$-values indicate that the polarized radiation on and off the MB probe distinctly different magnetic environments.

\subsection{Correcting for Faraday Rotation due to the MW Foreground}
\label{sec:MWcorr}

\begin{figure}
\centering
\includegraphics[width = 1.1\linewidth]{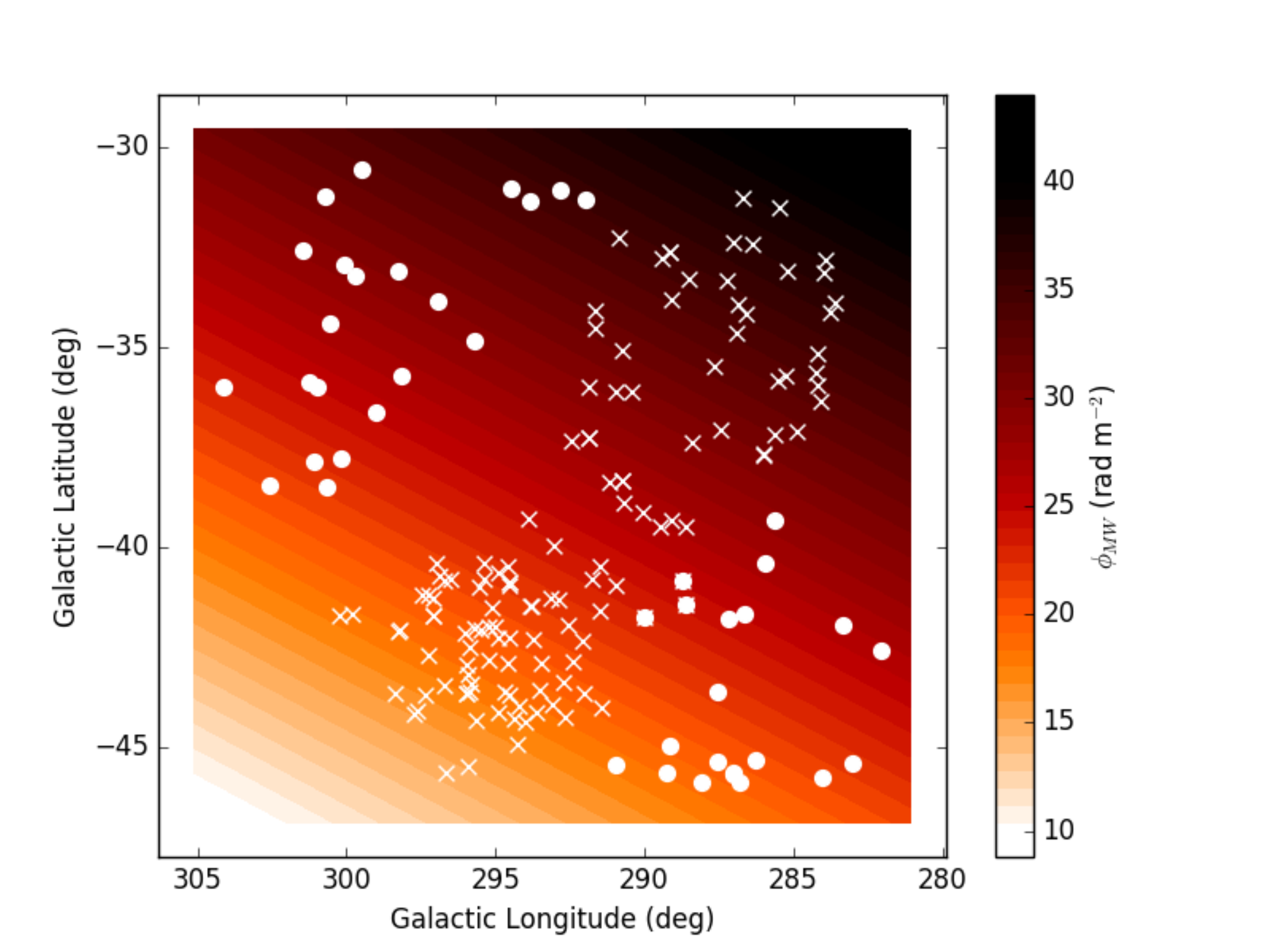}
\caption{An estimation of the foreground- and background-$\phi$ covering our field of view, assuming the Faraday depth varies as a tilted plane across our imaged region. The fit used the 43 off-Bridge sources which are shown as white dots. The location of the on-Bridge sources are shown as white crosses.}
\label{fig:rmSurface}
\end{figure}

The amount of Faraday rotation observed towards an extragalactic point source ($\phi_{\rm{obs}}$) will always include some contribution from the MW. Therefore, before the line-of-sight magnetic-field strength can be estimated, the Galaxy's contribution to the observed Faraday depth must be fit and corrected for. The 43 off-Bridge $\phi_{\rm{obs}}$ can be described by a tilted-plane $\phi_{\rm{MW}}$-model, whose parameters are obtained using a non-linear least-squares fit to the data. The best-fit solution was found to be of the form
\begin{equation}
	\phi_{\rm{MW}}\,=\,-0.511\ell\,+\,1.28{b}\,+\,225,
\end{equation} 
where $l$ and $b$ are the coordinates in Galactic longitude and latitude, respectively. The plane is shown in Figure\,\ref{fig:rmSurface}. By subtracting the resultant Faraday depth surface from all $\phi_{\rm{obs}}$, the residual Faraday depths ($\phi_{\rm{corr}}$ are considered to be foreground-corrected). 

We compare our MW Faraday depth model with similar models from \citet{Mao2008} and \citet{Oppermann2015}. Testing a point in the centre of the `Join' region ($\ell\,=\,290^{\circ}, b\,=\,-38^{\circ}$), our fit returns a $\phi_{\rm{MB}}$-value of $+28$\,rad\,m$^{-2}$. At the same position, MW models from \citet{Mao2008} and \citet{Oppermann2015} return values of $+28$\,rad\,m$^{-2}$ and $+25$\,rad\,m$^{-2}$, respectively. The close agreement amongst all three MW models adds confidence to our MW correction.

We further test the validity of the foreground $\phi_{\rm{MW}}$-model by comparing the distributions of the uncorrected and corrected $\phi$-values ($\phi_{\rm{obs}}$ and $\phi_{\rm{MB}}$, respectively) for points in the `North' and `South' regions. If the assumptions made to create the foreground model were valid, the distribution of Faraday depths should become more similar after the foreground correction has been applied. We test this theory by conducting two separate Anderson-Darling tests on the $\phi_{\rm{obs}}$ and $\phi_{\rm{corr}}$ distributions for the two off-Bridge regions. We find that before the foreground correction is applied there is $\sim{98\%}$ confidence that the two background samples are drawn from different populations. Once our model is subtracted from the raw, observed Faraday depths, the likelihood that the two populations are unique drops to $67\%$. At this level there is no longer sufficient confidence to say they are not drawn from the same parent distribution. We therefore consider our simplified tilted-plane assumption of the Faraday depth distribution of the MW-foreground to be justifiable.

Figure\,\ref{fig:corrRMs} shows the foreground-subtracted Faraday depths across the imaged region. We expect that after our foreground correction, the majority of off-Bridge sources would have values near zero, but this is not observed. We assume that the major cause for this discrepancy is that our foreground model is an oversimplification of the likely complex Faraday structure of the MW \citep{Oppermann2015}. We test the merit of a higher-order foreground Faraday depth model, but it produces minimal improvement while increasing the degrees of freedom. If our foreground fit was well founded, we would expect to have a mean $\phi_{\text{corr}}$-value of off-Bridge points near zero: our sample returns $\overline{\phi}_{\text{off, corr}}\,=\,0.3\,$rad\,m$^{-2}$ with a standard deviation of $12.0\,$rad\,m$^{-2}$, compared to $\overline{\phi}_{\text{off, obs}}\,=\,25\,$rad\,m$^{-2}$ before subtracting the foreground. We note that the foreground $\phi_{\rm{MW}}$ fit does not attempt to fit and subtract the Faraday rotation that is intrinsic to the background source. \citet{Schnitzeler2010} estimates the spread in intrinsic Faraday depths of extragalactic sources to be $\simeq\,6\,$rad\,m$^{-2}$, which can account for much of the large standard deviation of the off-Bridge, foreground-corrected Faraday depths. 

The uncertainty in the foreground Faraday depth subtraction must be included the error in the Faraday depth of the on-Bridge sources. The magnitude of the increased error was determined through bootstrapping the foreground $\phi_{\rm{MW}}$ surface 10,000 times with the standard deviation of the correction at each location ($\sigma_{\phi_{\text{MW}}}$). The mean uncertainty in Faraday depths through this method is $\overline{\sigma}_{\phi_{\text{MW}}}\,=\,0.21\,$rad\,m$^{-2}$. The expression for the total uncertainty in the Faraday depth of a background radio source therefore becomes
\begin{equation}
	d\phi(l,b)^{2}\,=\,d\phi_{\rm{MCMC}}^{2}\,+\,\sigma^{2}_{\phi_{\rm{MW}}}(l,b),
\end{equation}
where $(l, b)$ are the coordinates of the point source.

 We infer that the MB Faraday rotation, $\phi_{\text{MB}}$, accounts for the majority of the residual rotation seen in points associated with the MB and assume for all further analysis that $(\phi_{\text{obs}}\,-\,\phi_{\text{MW}})\,=\,\phi_{\text{corr}}$, where $\phi_{\text{corr}}\approx\,\phi_{\text{MB}}$. A map of foreground-corrected $\phi_{\text{MB}}$ is given in Figure\,\ref{fig:corrRMs}, which shows negative Faraday depths spanning the entirety of the MB. Analysis of this trend shows that $68\%$ of the polarized sources follow this trend to $2\times{d}\phi$, where $d\phi$ is the calculated error in our Faraday depth measurement.

$\phi_{\rm{MB}}$ may contain contributions from localised enhancements -- such as H\textsc{ii} and star formation regions -- that may influence the observed magnetic field on scales to which we are sensitive ($\sim{2}$\,pc). In order to identify any phenomena that could influence the small-scale magnetic field fluctuations in the MB, we cross-reference our region of sky with Simbad \citep{simbad} and find 7 molecular clouds \citep{Chen2014} and 4 H\textsc{ii} regions \citep{Meaburn1986, Bica2008} that are located in the `Wing' region. Three of the molecular clouds and three H\textsc{ii} regions are near the small patch of positive $\phi$-values near $l\,=\,295^{\circ}, b\,=-42^{\circ}$. These individual molecular clouds do not directly align with any of the background sources at our physical-scale sensitivity of $8\,$arcseconds.

\section{The Line-of-Sight magnetic-field strength}
\label{sec:B_strengths}

\subsection{Emission Measures}
\label{sec:Ha_and_EM}
Our objective is to calculate the line-of-sight magnetic field ($B_{\parallel}$) associated with the MB; however, $B_{\parallel}$ is degenerate with estimates of electron density ($n_e$). Therefore an independent estimate of $n_e$ is required. By making some assumptions about the line-of-sight depth of the ionized medium, it is possible to use observed H$\alpha$ intensities as a means to independently estimate $n_e^2$ by taking advantage of the implied emission measure (EM). The EM is defined as the integral of the square of the electron density along the pathlength of ionized gas ($L_{H\textsc{ii}}$) and can be derived from the measured H$\alpha$ intensity ($I_{\text{H}\alpha}$) in rayleighs ($R$)\footnotemark
\begin{equation}
\text{EM} = \int_{0}^{L}n_{e}(l)^{2}dl = 2.75\,T_4^{0.92}I_{\text{H}\alpha}~~\text{pc\,cm}^{-6}.
\label{eq:EM_dl}
\end{equation}

\begin{figure*}
\includegraphics[width =\linewidth,trim = {20 50 0 20}, clip]{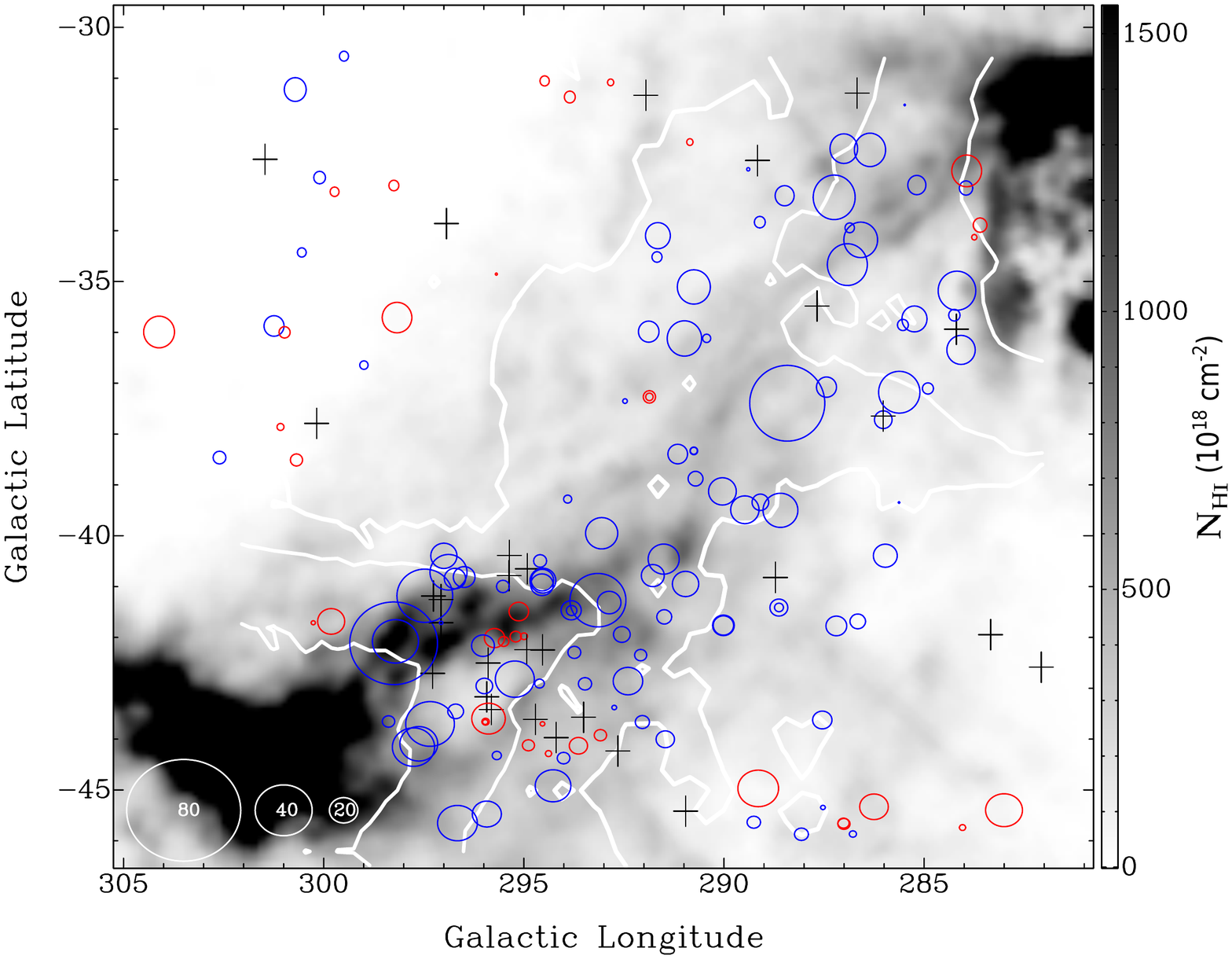}
\caption{Neutral Hydrogen intensity from GASS \citep{GASSIII} in the velocity range of $+100\,\leq\,v_{\rm{LSR}}\,\leq\,+300$\,km\,s$^{-1}$ overlaid with white contours representing non-extinction corrected H$\alpha$ intensities of 0.06, 0.15 and 1.0 $R$ as measured from WHAM \citep{BargerWHAM}. Circles represent the foreground-corrected Faraday depth ($\phi_{\rm{MB}}$) values towards each polarized background source. Red and blue circles represent a line-of-sight magnetic field oriented towards and away from the observer, respectively. Black crosses mark the existence of $\phi$ values that are consistent with zero to 2$\times{d}\phi$.}
\label{fig:corrRMs}
\end{figure*}

\footnotetext{$1\,\text{R}\,=\,(10^{6}/4\pi$)\,photons\,cm$^{-2}$\,s$^{-1}$\,s$r^{-1}$ which is equivalent to $5.7\times 10^{-18}$erg\,cm$^{-2}$s$^{-1}$\,arcsec$^{-2}$ for H$\alpha$.}
We utilise the work carried out by \citet{BargerWHAM}, which offers kinematically resolved intensities of the H$\alpha$ emission across our entire MB. The observations used in \citet{BargerWHAM} were made with the Wisconsin H$\alpha$ Mapper (WHAM) telescope, which has sensitivities of a few hundredths of a rayleigh (see \citealt{WHAM} for a complete summary of the telescope and survey technique). WHAM has a $1^{\circ}$ beam, which is equivalent to a diameter of nearly 1\,kpc at the assumed average distance to the MB of 55\,kpc. While the WHAM beam is considerably larger than the final resolution of our radio data, at this size it is less sensitive to small-scale H$\alpha$ emission stemming from individual H\textsc{ii} regions and is optimised to detect faint emission from diffuse ionized gas. For simplicity, we assume an electron temperature of $T_e = 10^4$\,K (denoted $T_4$), as assumed in \citet{BargerWHAM}. Figure\,\ref{fig:corrRMs} shows the MB region with white contours indicating levels of uncorrected H$\alpha$ emission from \citet{BargerWHAM}, tracing the 0.06, 0.15 and 1.0\,$R$ intensity levels.

Observed H$\alpha$ intensities are reduced from their intrinsic values due to dust contained within the MB itself and in the MW. These are known as internal and foreground extinction, respectively. We have corrected for both sources of extinction according to Table $2$ from \citet{BargerWHAM}. We assume that the `Join' and `West' regions have similar interstellar- and local dust content -- and therefore an identical total-extinction correction of $28\%$ has been applied. H$\alpha$-intensity correction of $22\%$ has been applied to all `Wing' points. For all future analysis and discussion, H$\alpha$ intensities have been extinction corrected, unless stated otherwise.

We cross-reference the position of each background polarized source with the WHAM data and accept the pointing with the smallest angular separation from our target as the representative H$\alpha$ brightness for that particular sightline. Because the WHAM survey of the MB is Nyquist sampled, the maximum angular separation allowed is less than 30\,arcminutes, which corresponds to $\leq\,500$\,pc at our assumed distance to the MB of 55\,kpc. EMs are then derived towards each matched sightline. Mean EMs for each region are listed in Table\,\ref{table:Bvalues}. 

\subsection{Distribution of Ionized Medium}
\label{sec:models}

In order to estimate the magnetic-field strength along the line-of-sight through the MB, we assume that there is no correlation between electron density and magnetic-field strength. This has been shown to be a reasonable approximation for typical gas densities associated with the diffuse interstellar medium \citep{Crutcher2003}. Rearranging Equation\,\ref{eq:RM}, it can be shown that the equation for magnetic field along the line-of-sight becomes
\begin{equation}
	B_{\parallel}\,=\,\frac{\phi_{\rm{MB}}}{0.812\,\overline{n_e}\,L_{H\textsc{ii}}},
	\label{eq:Blos}
\end{equation}
 where $\phi_{\rm{MB}}$ is the MW-foreground corrected Faraday depth and $\overline{n_e}$ is the mean electron density along the total pathlength of ionized material ($L_{H\textsc{ii}}$).
  
Often, pulsar dispersion measures (DM\,$=\,\overline{n_e}L_{H\textsc{ii}}$) can be used to construct well-formed estimates of the pathlength and electron density through the different regions. Unfortunately, there are no known pulsars in the MB and very little is known about the morphology and line-of-sight depth of the MB. 

\citet{Subramanian2009} argue that the SMC is nearly edge-on, indicating a pathlength through the galaxy of $\geq\,5\,$kpc. If the bulk of the material in the MB had its origins in the SMC, one might expect the depth of the MB to be equally large. \citet{Muller2004} argue that there are numerous observations throughout the MB that hint at a large line-of-sight depth and \citet{Gardiner1994} estimate the pathlength through regions of the MB 5\,kpc\,$\lesssim\,L\,\lesssim$\,10\,kpc. For simplicity, we parameterise and evaluate the depth of the MB as $L_5\,=\,5$\,kpc and consider the implications of different pathlengths through this parameter, with $1\lesssim{L_5}\lesssim{2}$.

Several independent assumptions corresponding to the distribution and geometry of ionized- and neutral-gas can be made in order to validate our $B_{\parallel}$ measurements. Below, we describe three separate ionized gas distributions and discuss how each might affect derived magnetic-field strengths. In our discussion, all ionized parameters will be denoted with subscript H\textsc{ii} and all neutral gas parameters will be denoted with subscript H\textsc{i}, unless otherwise noted. 

\subsubsection{Case 1: Constant Dispersion Measure}
\label{sec:Bmods_1}

When estimating the line-of-sight magnetic field strength, the simplest model of the distribution of material in the MB is one in which the neutral and ionized gas are well-mixed. In such a scenario, the bulk of the neutral gas would be distributed across the MB in small clumps, with the ionized medium distributed uniformly amongst the neutral clouds. Therefore, the effective depth of the ionized medium can be expressed as a fraction of the depth of the neutral material, $L_{H\textsc{ii}}\,=\,fL_{\text{H}\textsc{i}}$, where $L_{\text{H\textsc{i}}}$ is the depth of the neutral gas and $f$ is the filling factor of ionized gas along the total line-of-sight \citep{Reynolds1991}. 

Little is known of the effective filling factor of ionized gas along the line-of-sight, but a filling factor of $f\,=\,1$ is highly unlikely. Previous work on nearby high-velocity clouds in the Leading Arm \citep{McClureGriffiths-LA-2010} has assumed a filling factor of $f\,\sim\,0.5$ to describe the distribution of the ionized gas and we assume the same value for our analysis. In Section\,\ref{sec:Bmods_comp}, we briefly explore the implications of a range of filling factors. Combining the derived EM with our line-of-sight estimates, the DM becomes
\begin{equation}
	\text{DM}\,=\,(\text{EM}\,f\,L_{\text{H}\textsc{i}})^{1/2}.
\end{equation}

Incorporating the above expression for DM with Equation \ref{eq:Blos}, estimates of the magnetic field along the line-of-sight can be evaluated as
\begin{equation}
		B_{\parallel}\,=\,\frac{\phi_{\rm{MB}}}{0.812\,(\text{EM}\,f\,L_{\text{H}\textsc{i}})^{1/2}}.
	\label{eq:Blos_mixed}
\end{equation}
 This assumption of the geometry of the ionized material in the MB is likely an oversimplification of the actual distribution, which is expected to vary as a function of position along the MB.
 
\subsubsection{Case 2: Constant Regional Ionization Fraction}
\label{sec:Bmods_2}
In contrast to Case 1, where we estimated the effective pathlength of the ionized material, here we estimate the free-electron content of a sightline using the ionization fractions ($X$) across the MB. In order to motivate this approach, a few assumptions must be made. Firstly, we assume that the bulk of the MB material is in the velocity range $+100\,\leq\,v_{\rm{LSR}}\,\leq\,+300$\,km\,s$^{-1}$ relative to the Galactic centre \citep{Putman2003, Muller2003}. 

Following from the previous assumption, we also assume that the observed H\textsc{i} depth from GASS \citep{GASSIII} in our selected velocity range probes the entire line-of-sight depth of the MB such that the ionization fraction of a region represents the sum of ionized material in the MB along a given sightline. Previous investigations into the MB have shown this assumption to be reasonable in the diffuse regions of the MB, where observation have shown there to be little dust content \citep{Smoker2000, Lehner2008}. However there have been observations of molecules in the `Wing' region \citep{Muller2004, Mizuno2006, Lehner2008} and this assumption will serve as a lower limit to our estimates of neutral- and ionized-gas densities in this region. This second assumption indirectly implies that the neutral and ionized gas are well mixed (i.e. $f\sim1; L_{\text{H}\textsc{i}}\simeq{L}_{\text{H}\textsc{ii}}$) since any reported ionization fraction is reflective of the pathlength of neutral gas. 

Following from these assumptions, the electron density is calculated simply as the ionization fraction multiplied by the neutral-gas density
\begin{equation}
\overline{n_{e}} =\frac{X\,\langle{N_{\text{H}\textsc{i}}}\rangle}{f\,L_{\text{H}\textsc{i}}}.
\label{eq:n_0}
\end{equation}
As with Case 1, the above expression has the underlying premise of $L_{\text{H}\textsc{ii}}\,=\,f\,L_{\text{H}\textsc{i}}$. It follows then that the DM can be written as
\begin{equation}
	\text{DM}\,=\,\frac{X\,\langle{N}_{\text{H}\textsc{i}}\rangle}{f\,L_{\text{H}\textsc{i}}}{L_{\text{H}\textsc{ii}}}\,=\,3.09\times{10^{18}\,}X\langle{N}_{\text{H}\textsc{i}}\rangle,
	\label{eq:DM}
\end{equation}
where the constant $3.09\times{10^{18}}$ is the conversion factor of pc to cm.

With an expression for the DM, it is now possible to estimate the magnetic field along the line-of-sight by combining Equations\,\ref{eq:Blos} and \ref{eq:DM},
\begin{equation}
	B_{\parallel}\,=\,3.80\times{10^{18}}\left(\frac{\phi_{\rm{MB}}}{X\,\langle{N_{\text{H}\textsc{i}}}\rangle}\right).
	\label{eq:ionFrac}
\end{equation}

The MB is highly ionized with ionization fractions dependent upon location within the MB \citep{Lehner2008, BargerWHAM}. \citet{BargerWHAM} determined the minimum multiphase ionization fraction across the MB and argued that in the case where the neutral and ionized gas is well-mixed, the average ionization fraction in the region of the diffuse MB is $X\simeq46\%$, and  $X\simeq29\%$ in the `Wing' region. As these ionization fractions represent the average values calculated over the entire region, the values do not represent small-scale variations in the distribution of material. In contrast, the spectroscopic work of \citet{Lehner2008} found ionization fractions as high as $X\simeq90\%$ along three sightlines corresponding to the `Join' and `West' regions. As their sightlines probed localized distributions, this approach would have been susceptible to small-scale enhancements. 

Motivated by the wide variability in ionization fractions, we choose to evaluate the ionization level of the various regions individually. In the region of the Wing, we compare the H\textsc{i} and H$\alpha$ column densities from Table 3 in \citet{BargerWHAM}, to calculate a multiphase ionization fraction of $X\simeq29\%$ in the `Wing'. We evaluate the `Join' and `West' regions at an ionization fraction of $X\simeq46\%$, assuming that the `Join' and `West' regions host similar distributions of material. Evaluation of this ionization level makes it simple to explore the range of possible magnetic field strengths.

In the region of the SMC-Wing, there is a clear variation of H\textsc{i} column densities as well as H$\alpha$ intensities (Figure\,\ref{fig:corrRMs}). Following \citet{BargerWHAM}, we choose to break up the Wing into two regions corresponding to the relative H$\alpha$ brightness. If a sightline is associated with an uncorrected H$\alpha$ brightness larger than 0.15\,R, this region is classified as the `H$\alpha$-Wing' and assigned a H\textsc{i} column density of $6.8\times10^{20}\,$cm$^{-2}$, else we consider the region to be the `H\textsc{i}-Wing' and evaluate it as having a H\textsc{i} column density of $3.6\times10^{20}\,$cm$^{-2}$. Furthermore, assuming well-mixed neutral and ionized gas populations (e.g. $L_{\text{H}\textsc{I}}\,\simeq\,L_{\text{H}\textsc{II}}$), we use the mass estimates of \citet{BargerWHAM} to calculate an ionization fraction of $X\simeq24$ and $X\simeq21$ for the `H\textsc{i}-' and `H$\alpha$-Wing', respectively. A summary of region parameters is given in Table\,\ref{table:Bvalues}.

This assumed geometry of the distribution of ionized gas is similar to Case 1 ($\S$5.2.1), in that it requires the neutral and ionized media to be well-mixed. However, in this model, our greatest approximation is the mean ionization fraction for a given region of the MB. Although not stated explicitly in Equation\,\ref{eq:ionFrac}, this B$_{\parallel}$ estimate does have a dependence on the assumed pathlength of ionized material through the derivative of the total ionized mass of the region and subsequently implied $X$, which is outlined in \citet{BargerWHAM}.%
\subsubsection{Case 3: Ionized Skin}
\label{sec:Bmods_3}

Ionising photons that have escaped the MW and the Magellanic Clouds have the potential to ionise the outer layers of the MB \citep{Fox2005, BargerWHAM}. In this possibility, the distribution of the thermal electrons is that of an ionized skin, rather than mixed with the neutral gas, as we assumed in Cases 1 and 2. In order to explore this third scenario, we assume that the neutral hydrogen is girt by a fully ionized skin at the same temperature and pressure, the density of which will be $\overline{n_e}\,=\,\overline{n_{\text{H}\textsc{i}}}/2$ \citep{Hill2009}. This condition requires that the neutral and ionized media have had enough time to come into pressure equilibrium, which we assume for our analysis.

In the ionized skin, the line-of-sight depth can be derived from our density assumption combined with H$\alpha$ brightnesses:
\begin{equation} 
L_{H\textsc{ii}}\,=\,\text{EM}\,n_{e}^{-2}\,=\,\text{EM}\,\left(\frac{\overline{n_{\text{H}\textsc{i}}}}{2}\right)^{-2}\,=\,4\,\text{EM}\,\left(\frac{f\,L_{\text{H}\textsc{i}}}{\langle{N_{\text{H}\textsc{i}}}\rangle}\right)^{2},
\label{eq:L_H+}
\end{equation}
where the discussion for the evaluation of $n_e$ is given in the previous model (Equation \ref{eq:n_0}). We can now combine Equation \ref{eq:L_H+} with our density estimates (Equation \ref{eq:n_0}) to find an expression for DM:
\begin{equation}
	\text{DM}\,=\,\frac{2\,\text{EM}\,f^2\,L_{\text{H}\textsc{i}}}{\langle{N_{\text{H}\textsc{i}}}\rangle}.
	\label{eq:DM_skin}
\end{equation}

\noindent Substituting this expression for DM into Equation\,\ref{eq:Blos}, the equation for the magnetic-field strength along the line-of-sight in an ionized skin becomes

\begin{equation}
	B_{\parallel}\,=\,\frac{\phi_{\rm{MB}}\,\langle{N_{\text{H}\textsc{i}}}\rangle}{1.6\,\text{EM}\,f^2\,L_{\text{H}\textsc{i}}}
	\label{eq:B_ionSkin}
\end{equation}
where the line-of-sight of the neutral medium is in units of cm.

In the case of an ionized skin, the pathlength of the ionized medium is expressed explicitly in terms of our two assumptions: firstly, that the neutral and ionized media are in pressure equilibrium and secondly, that the filling factor of the neutral medium is $f_{\text{H}{\textsc{i}}}\,\simeq\,1$ along the effective depth of the MB. Therefore, we argue that for the above thin-skin approximation, $f\,\simeq\,1$. 

\subsection{Summary and Comparison of Ionization Cases}
\label{sec:Bmods_comp}

\begin{figure}
\includegraphics[width = 1\linewidth]{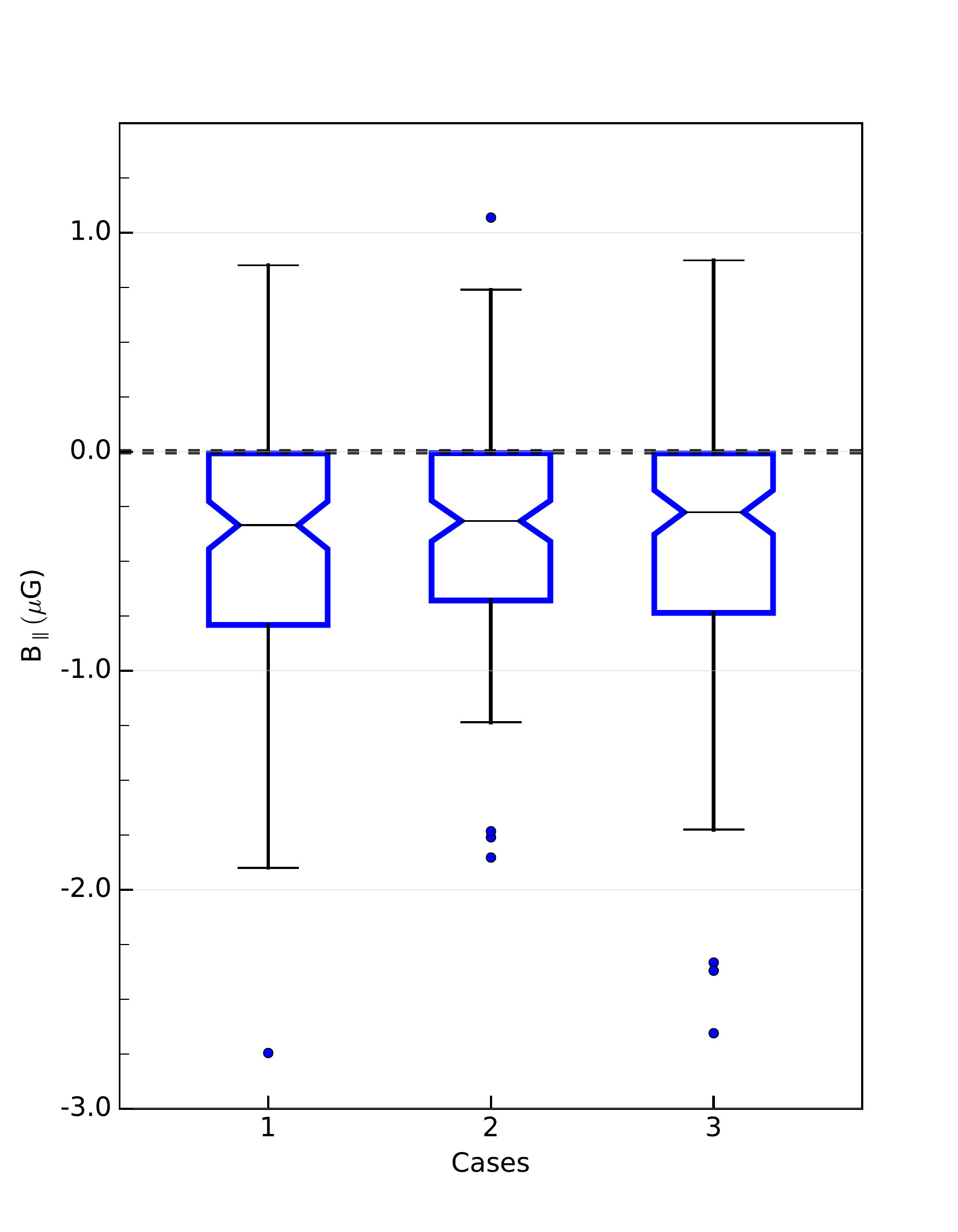}	
\caption{Box-and-whisker plots of coherent line-of-sight magnetic field measurements towards all sightlines through the MB for all cases listed in $\S$\ref{sec:models}. The height of each box marks the IQR of each distribution while notches mark the median position. The caps at the end of the whiskers represent the 5th and 95th percentiles. The outliers in each population are shown as dots above and below the whiskers. There is a dashed line at $B_{\parallel}=0$ to clarify the distribution of positive and negative magnetic field orientations. }\label{fig:boxPlots}
\end{figure}

We evaluate each of the aforementioned cases for a line-of-sight pathlength of $L_{\text{H}\textsc{i}}\,=\,L_{5}\,=\,5\,$kpc. As shown in Table\,\ref{table:Bvalues} and Figure\,\ref{fig:boxPlots}, each of the cases results in similar estimates in line-of-sight magnetic-field strengths for the entire MB, with median values all near $B_{\parallel}\,\simeq\,0.3\,\mu$G. By comparison, individual regions show a larger scatter between derived magnetic-field strengths. 

Our sample of Faraday depths is skewed towards the negative (as seen in Figure\,\ref{fig:corrRMs}); therefore, it follows that the derived field strengths are distributed in kind. By completing a skewness test, we find that the $B_{\parallel}$ distribution resulting from Case 1 is skewed towards negative values with a $3.2\,\sigma$ confidence level. It follows that Case 2 is skewed negative to $1.7\sigma$ and Case 3 to $4.2\sigma$ significance. This skew can be seen most clearly in Figure\,\ref{fig:boxPlots}. Due to this skew towards negative values, the $B_{\parallel}$ values quoted in Table\,\ref{table:Bvalues} represent the median magnetic-field strengths, where the median statistic is more robust against outliers. Along with the median value, we list the deviation from the first and third quartile ($\mathcal{Q}_1$ and $\mathcal{Q}_3$, respectively), which represents the 25th and 75th percentile values in the distribution. The derived magnetic-field strengths are best summarised by Figure\,\ref{fig:boxPlots}, where the bound region denotes the interquartile range (IQR), defined as IQR\,=\,$\mathcal{Q}_3\,-\,\mathcal{Q}_1$. 

As noted in our discussion of the ionization models, the largest uncertainty in our $B_{\parallel}$ measurements comes from the unknown geometry of the MB along the line-of-sight, namely the uncertainty in $L_{\text{H}^+}, X$, and $f$. With that in mind, we aim to compare all models by their dependence on our depth assumptions and use of measured quantities. 

Cases 1 and 2 are built from the same oversimplified picture of well-mixed neutral and ionized gas distributions along the line-of-sight. Case 1 uses only the measured EM with a largely unconstrained filling factor, $f$. Exploring a range of $f$ for Case 1 shows that a $20\%$ change in $f$ (i.e. $0.3\leq\,f\,\leq0.7$) results in less than a $0.1\,\mu$G change in the median line-of-sight magnetic field strength. In contrast, Case 2 takes advantage of more information, using both calculated ionization fractions and measured $\langle{N_{\text{H}\textsc{i}}}\rangle$. Contrasting these first two ionized gas distributions, Case 3 has the ionized material distributed as an ionized skin. This geometry requires that the neutral and ionized gas to be in pressure equilibrium in order to be physical. It is possible that this condition could be met in the `Join' region; however, it is likely that this is inappropriate for regions in the `Wing' due energy being injected from on-going star formation (e.g. \citealt{Noel2015}).

We show that Case 2 has least dependence on an assumed pathlength through the MB and filling factor. As we mentioned in $\S$\ref{sec:Bmods_2}, the actual ionization fractions across the MB may be higher than our evaluated estimates. Increasing the ionization fraction by a factor of two implies that the line-of-sight magnetic field strength is half the current value -- i.e. $B_{\parallel}(X=90\%)\,\simeq\,-0.17\,\mu$G. The following discussion will be carried out using the magnetic field estimates derived from Case 2 and all parameters reported in Table\,\ref{table:Bvalues}, unless specified otherwise.

\begin{savenotes}
\begin{table*}
\begin{tabular}{l r c c c c c c | c c c | c}
\multicolumn{3}{c}{(1)} 	&		(2)	 &		(3)		 &	(4)		&		(5)		 & 	(6)		&	 	(7)		&		(8)		&		(9)		&		(10)			\\[3.5pt]
\multicolumn{3}{c}{Region} 		&	$\langle{N_{\text{H}\textsc{i}}}\rangle$	 &	 $\langle$EM$\rangle$\footnotemark	&		 $\overline{\phi}_{MB}$		&		$\sigma(\phi)$		& 	$X$		 & 
																			$B^{*}_{\parallel,1}$	&	 $B^{*}_{\parallel,2}$	&	$B^{*}_{\parallel,3}$		&		$B_r$	\\[2pt]
					&&&($\times{10^{20}}\,$cm$^{-2}$) &	(pc\,cm$^{-6}$)		&		(rad\,m$^{-2}$)&	(rad\,m$^{-2}$)	& 	(\%)	&	($\mu$G\,$L_{5}^{-1/2}$)	&	($\mu$G)	&		($\mu$G\,$L_{5}^{-1}$)		&	($\mu$G)\\[2pt]
					\hline\hline
&&&&&&&&&\\[-2mm]
\multicolumn{3}{c}{Join}	&	$3.0$		&		$0.283$		&		$-13.5$	&		$9.3$			&		$46$	&		$-0.61^{+0.22}_{-0.41}$		&		$-0.43^{+0.19}_{-0.15}$		&		$-0.56^{+0.28}_{-0.57}$		&	 $0.11$	\\[5pt]
\multicolumn{3}{c}{West}&	$2.8$		&		$0.603$		&		$-12.3$	&		$15.4$		&		$46$	&		$-0.33^{+0.29}_{-0.44}$		&		$-0.27^{+0.23}_{-0.41}$		&		$-0.23^{+0.20}_{-0.25}$		&	$0.84$	\\[5pt]
\multicolumn{3}{c}{Wing}&	$5.0$		&		$0.823$		&		$-8.7$	&		$15.4$		&		$29$	&		$-0.26^{+0.32}_{-0.45}$		&		$-0.29^{+0.35}_{-0.43}$		&		$-0.27^{+0.21}_{-0.57}$		&	$1.0$		\\[5pt]
&	&	$-$\,H\textsc{i} Wing	&	$3.6$		&		$0.295$		&		$-8.1$	&		$12.5$		&		$24$	&		$-0.39^{+0.44}_{-0.40}$		&		$-0.40^{+0.46}_{-0.35}$		&		$-0.44^{+0.48}_{-0.50}$		&		$$		\\[5pt]
&	&	$-$\,H$\alpha$ Wing	&	$6.8$		&		$1.69$		&		$-9.7$	&		$19.1$		&		$21$	&		$-0.06^{+0.14}_{-0.32}$		&		$-0.07^{+0.16}_{-0.38}$		&		$-0.08^{+0.17}_{-0.36}$		&		$$		\\[5pt]
\hline
\multicolumn{3}{c}{\bf{Total}}&		&							&						&						&						&		$-0.34^{+0.33}_{-0.45}$		&		$-0.32^{+0.31}_{-0.36}$		&		$-0.28^{+0.27}_{-0.46}$		&					\\[5pt]
	\end{tabular}
	\caption{Table of derived values for polarized sources in all regions of the MB. Column (1) specifies the region of interest, column (2) gives the average H\textsc{i} column density for the region as measured from GASS \citep{GASSIII} and column (3) gives the average extinction-corrected EM from the WHAM dataset \citep{BargerWHAM}. The mean foreground-corrected Faraday depth is given in column (4). Column (5) gives the standard deviation of the foreground-corrected Faraday depth of the region about the mean. The average ionization fraction for each region, assuming the neutral and ionized material is well mixed, as determined by \citep{BargerWHAM}, is listed in column (6). Columns (7\,-\,9) give the median coherent magnetic-field strength along the line-of-sight for each of the ionization geometries discussed in Section $\S$\ref{sec:models}. The errors listed represent the deviation from the 25th and 75th percentiles. The implied random magnetic-field strength, as calculated from Equation\,\ref{eq:B_rand} is given in column (10). }
\label{table:Bvalues}
\end{table*}
\end{savenotes}

\section{Discussion}
\label{sec:discuss}

In this section, we consider the implications of the observed Faraday-depth values and magnetic-field strengths in the MB and explore possible origins of the coherent magnetic structure.

\subsection{The Turbulent Magnetic Field}
\label{sec:Turb_bField}

On-going star-formation in the MB (e.g. \citealt{Noel2015}) will make any existent regular magnetic field to become turbulent and random. An increase in random motion would also depolarize any background polarized light, proportional to the level of turbulence. If the magnetic field observed in the MB were sufficiently turbulent, one would expect that the polarization of sources associated with the MB would exhibit higher levels of depolarization, and thus have lower values for the observed fractional polarization. 

\begin{figure}
\includegraphics[width=1.\linewidth]{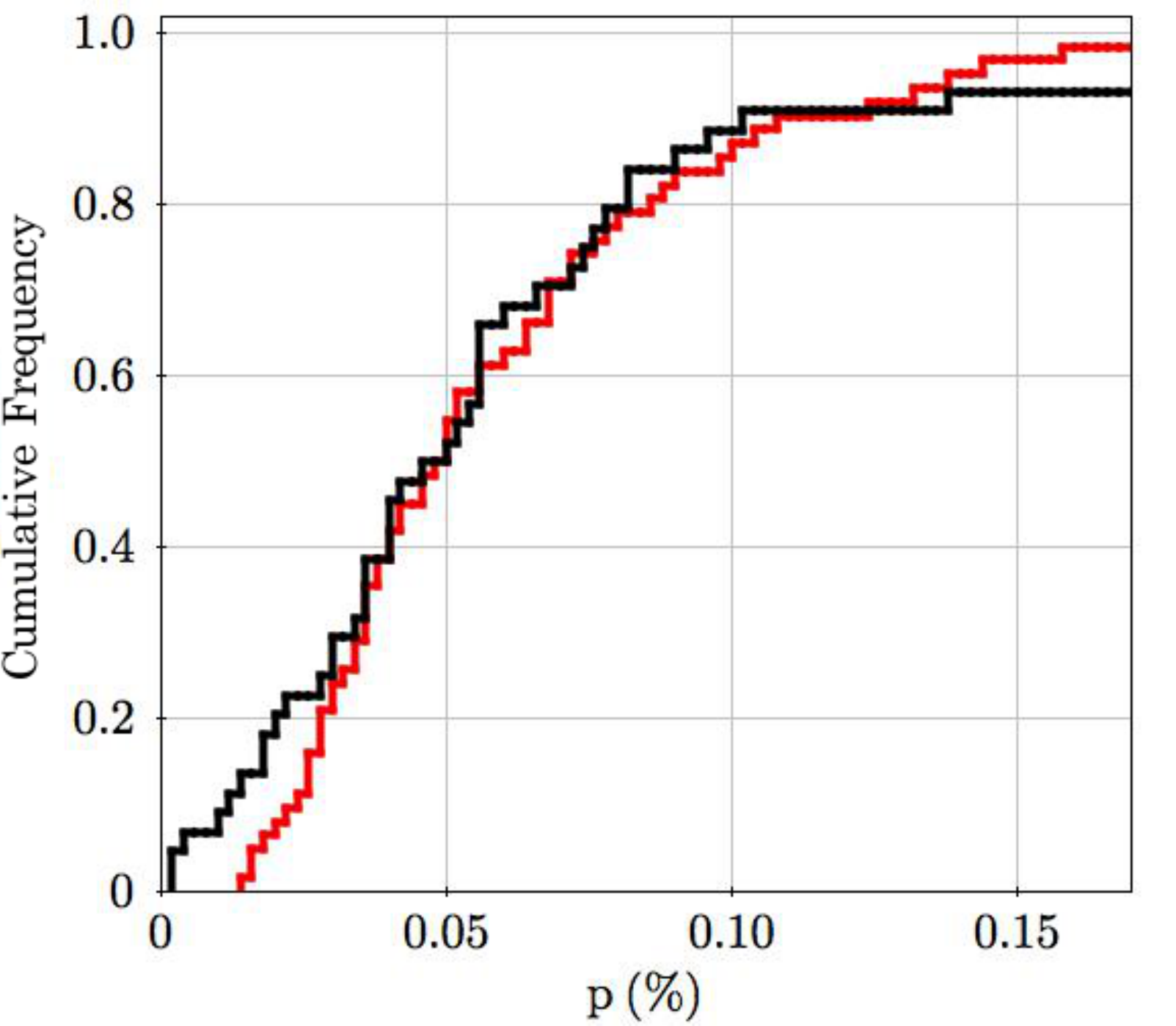}
\caption{Cumulative histogram comparing observed polarization fractions ($p$) for sources on (red) and off the Bridge (black). Sources associated with the `Wing' region are not included in this distribution due to the source selection bias towards sources with high polarization. This figure has been truncated at $p\,=\,17\%$ for clarity.}
\label{fig:cumHist_polfrac}
\end{figure}

We explore the consequences of the turbulent field by comparing the observed polarization fraction for populations of sources on and off the MB. Figure\,\ref{fig:cumHist_polfrac} shows a cumulative histogram comparing the observed polarization fractions. We choose not to include sources associated with the `Wing' region due to the source selection bias that favours highly-polarized sources. The two source populations show no statistically-significant differences in the observed fractional polarization, indicating there is no correlation between source location and turbulence of the foreground magnetic field.

\begin{figure}
\includegraphics[width = 1\linewidth]{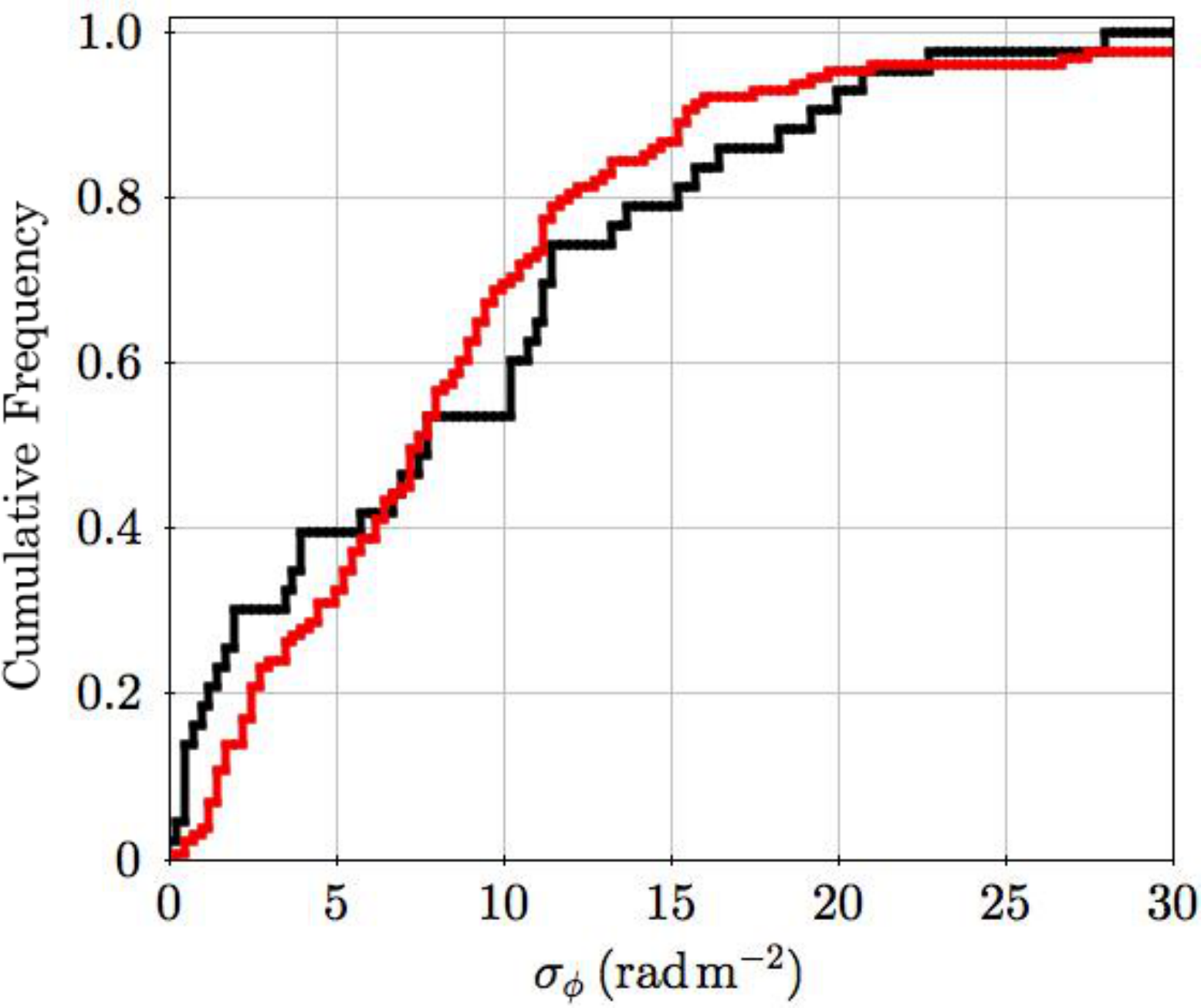}
\caption{Cumulative histogram of best-fit Faraday-dispersion values ($\sigma_{\phi}$) fitting to a single, simple Faraday-rotating source with foreground depolarization for sources on (red) and off the Bridge (black). The figure has been truncated at $\sigma_\phi\,=\,30\,$rad\,m$^{-2}$ for clarity.}
\label{fig:cumHist_sigma}
\end{figure}

If the turbulence in the field is not strong enough to depolarize the background signal completely, it is still possible to investigate the mean Faraday dispersion ($\sigma_{\phi}$) as fitted by our $qu$-fitting routine. We compare the values for sources on-Bridge and off-Bridge, under the hypothesis that sources on the MB would exhibit higher $\sigma_{\phi}$ if there are more coherent and/or turbulent cells located in the MB when compared to the MW. Figure\,\ref{fig:cumHist_sigma} shows a cumulative histogram of the best-fit Faraday dispersion values for all points on and off the MB. We carry out a two-sample Anderson-Darling test with both $\sigma_{\phi}$ populations and find that we cannot reject the null-hypothesis of the two samples being drawn from the same distribution and conclude that any turbulence in the MB magnetic field cannot be differentiated from that in the MW. 

In the MW, LMC and SMC, it has been shown that the random component of the magnetic field dominates the total field strength (\citealt{Beck2000, Gaensler2005, Mao2008}). To estimate the random magnetic-field strength, we choose a similar approach to \citet{Mao2008}, who assume that the coherent magnetic field does not change as a function of position, but that any change in observed $\phi$ is due to turbulence. It is then possible to estimate the mean random magnetic-field strength for each observed region as

\begin{equation}
B_r\,=\,\frac{3l_0}{L_{5}}\,\sqrt{\left(\frac{\sigma(\phi)}{0.812\,\overline{n_e}\,l_0}\right)^2 -\left(\frac{B_{\parallel}^{*}\,\Delta{L}}{l_0}\right)^2}, 
\label{eq:B_rand}
\end{equation}

\footnotetext{The mean EM is not used in the derivation of magnetic-field strengths. It is listed to give the reader an appreciation of the characteristics of the region.}

\noindent where $\sigma(\phi)$ is the standard deviation of Faraday depths in the region of interest, $\overline{n_e}$ is the average electron density in units of cm$^{-3}$ and $B_{\parallel}^{*}$ is the median coherent magnetic-field strength along the line-of-sight in $\mu$G. $\Delta{L}$ is the standard deviation of the pathlength through the ionized medium in pcs and characterises the uncertainty in the depth of the MB. $l_0$ is the linear scale of a `RM-cell' in units of pc, such that $n\,\sim\,L_{5}/l_0$, where $n$ is the number of cells in a single line-of-sight.

As stressed in our derivation of the coherent magnetic-field strength, little is known about the morphology of the MB, leaving the estimations for pathlength to be our largest uncertainty. We estimate the standard deviation of the width of the MB ($\Delta{L}$) to be 1\,kpc. \citet{Gaensler2005} show that RM-cells in the LMC are of order $\sim100$\,pc, and we adopt a similar value for our analysis. As we have done in Case 2 ($\S$\ref{sec:Bmods_2}), we consider $\overline{n_e}$ to be related to the column density of H\textsc{i} as $\overline{n_e}\,=(X\,\langle{N_{\text{H}{\textsc{i}}}}\rangle)/L_{5}$ where $L_{5}$ is the line-of-sight depth of the MB in cm and $X$ is the ionization fraction of the region. 

As discussed in Section\,$\S$\ref{sec:MWcorr}, our correction for the foreground MW Faraday depth does not account for the intrinsic Faraday depth of the source. We minimise any resultant effects by subtracting the scatter of intrinsic extragalactic Faraday depths $\sigma(\phi)=6$\,rad\,m$^{-2}$ \citep{Schnitzeler2010} from our regional Faraday depth standard deviations. 

Using the above estimates and the values listed in Table\,\ref{table:Bvalues} we derive the implied random magnetic-field strengths of each region, the results of which are summarised in Table\,\ref{table:Bvalues}. We find that the turbulent field dominates the ordered component in the regions of the `Wing' and `West'. Intriguingly, this does not hold in the `Join' region. Perhaps this is indicative that our pathlength estimates, are unrealistic or that our overarching assumptions are unviable. However, the `Join' region is furthest from any on-going star-formation. This fact, combined with with our aforementioned turbulence null-hypotheses, suggests that the random field may not dominate the large-scale magnetic field in the diffuse MB. 

\subsection{Estimating the Total Magnetic Field of the Wing}
\label{sec:wingB}

Recent work by \citet{LoboGomes2015} mapped coherent field lines in the plane-of-the-sky ($B_{c, \perp}$) in the SMC and Wing using optical polarized starlight. They argued that there exists a significant fraction of sightlines that exhibit a magnetic field that points in the direction of the MB towards the LMC of order $B_{\perp}\,=\,(0.947\,\pm\,0.079)\,\mu$G. 

We combine their measurements with our estimations for $B_{\parallel}$ to estimate the total coherent magnetic-field strength ($B_{c,\,T}$) in the Wing 
\begin{equation}
	B_{c,\,T}^{2}\,=\,B_{c,\,\perp}^{2}\,+\,B_{c,\,\parallel}^{2}.
\end{equation}
 We find an implied total magnetic-field strength of $B_{c,\,T}\,\simeq\,1\,\mu$G in the region of the Wing which implies that the ordered magnetic field in the Wing is dominated by the plane-of-the-sky component. We note that our uncertainty estimates imply a large range of coherent field strengths. For the sake of brevity, we omit the implications of all possible field strengths from our discussion. This field strength is within the range of magnitudes expected if the field were to have originated from the SMC as \citet{Mao2008} estimated total coherent magnetic field in the SMC to be $B_T\,\simeq\,1.7\,\pm\,0.4\,\mu$G. 
  

\subsection{The Pan-Magellanic Field}

The possible existence of a large-scale magnetic field that permeates the entire Magellanic System (the `pan-Magellanic field', pM field) was first introduced by \citet{MathewsonFord1970b} and \citet{Schmidt1970}. Furthermore, \citet{Schmidt1970} argued that the existence of such a field suggested that the fields observed in the LMC and SMC shared a common origin. Continued investigations into the nature of the magnetism across Magellanic System were carried out (e.g. \citealt{Schmidt1976, Mathewson1979, Wayte1990}) strengthening the case for the existence of the pM field. More recently, \citet{Mao2008} and \citet{LoboGomes2015} note the potential alignment of the SMC magnetic field with the MB, and \citet{Mao2012} argue the same for the LMC. However, all previous research has been confined to high density regions in the LMC and SMC. 

If the pM field exists, it is expected to be dominated by the plane-of-the-sky component, just as the fields associated with the LMC and SMC have been observed to be \citep{Mao2012, LoboGomes2015}. However, the observation of negative Faraday depths across the MB implies a non-trivial line-of-sight component. We argue that this directional component was anticipated, as the SMC is located further away from the MW than the LMC ($\sim60$ and $\sim50$\,kpc, respectively \citealt{Walker1999}). Therefore, our observation of a Faraday-depth signal spanning the entirety of the MB may be the first direct evidence of the pan-Magellanic field.

Before we can confirm the existence of the pan-Magellanic field, it is important to understand if the coherent fields associated with the SMC and LMC can account for the observed Faraday depth signal seen to span the entire MB. Below we investigate the possible origins of the observed coherent magnetic field and how it might relate to the `pan-Magellanic field' hypothesis. We assume that the observed magnetic field has been frozen in to the tidally-stripped gas for this discussion, and address any evidence towards the contrary at the end of this section.

\subsubsection{The SMC-Wing Field}
\label{sec:SMC-Wing}
\citet{Mao2008} estimate that the SMC has a line-of-sight and total magnetic-field strength of $B_{\parallel}\,=-0.19\pm\,0.06$ and $B_{Tot,C}\,=\,1.6\pm0.4\,\mu$G, respectively. All ionization models discussed in our paper imply a line-of-sight magnetic-field strength that is consistent with this estimate in the `Wing' region, within the IQR (Table\,\ref{table:Bvalues}). We note that our average median value between all Cases is $\overline{B}^*_\parallel\,=\,-0.27\,\mu$G, a value that is larger than what was observed in the SMC. However, our estimates are again based largely on the assumed geometry of the MB. If the MB is oriented predominantly along the line-of-sight, as is the orientation of the SMC, then the ionized pathlength would be larger than $L_5$. Doubling the line-of-sight depth through the Bridge ($2L_5$) results in a decreased estimated magnetic-field strength of $\overline{B}^*_\parallel\,=\,-0.20\,\mu$G, a value that agrees well with the magnetic field estimates of \citet{Mao2008}.

If the SMC magnetic field is responsible for the observed MB field, it is also possible to estimate the expected MB $\phi$ signal. Using Case 1 ($\S$\ref{sec:Bmods_1}) and an average pathlength of $\overline{L}\,=\,L_5\,$, a filling factor of $f\,=\,0.5$ and estimates of the average electron density from the EMs from Table\,\ref{table:Bvalues}, a coherent line-of-sight magnetic field of $B_{\parallel}$ would manifest itself as a Faraday depth of $\phi\,\sim\,-7$ and $-4$\,rad\,m$^{-2}$ in the `Wing' and `Join' regions, respectively. This approximation is roughly consistent with the observed median $\phi$ in the Wing, but appears to contradict the mean Faraday depth measured in the `Join' region, which returns an average Faraday depth that has a magnitude more than 3 times the expected value ($\overline{\phi}_{MB}\,=\,12.5\,$rad\,m$^{-2}$). To achieve the observed mean Faraday depth in the `Join' region with the SMC $\overline{B_{\parallel}}$ requires an effective pathlength of $\sim\,27\,$kpc, which is unlikely to be physical.

This contradiction can be accounted for if the orientation of the magnetic field changes as a function of position along the MB. If in the `Join' region, the coherent field has been rotated such that a larger fraction of the total field lies along the line-of-sight, one would observe a larger $\phi_{\rm{MW}}$ and derive a stronger $B_{\parallel}$. Indeed, this is what is observed in the `Join' region.

We argue that the magnetic field present in the `Wing' region is consistent with a field that was created in the SMC and pulled into the MB. It is possible that the inherited field stretches as far as the `Join' region. However, it requires that the turbulent field to be less significant than that associated with the `Wing' region. It is then plausible that the pan-Magellanic field is governed by the geometry of the coherent field in the SMC. We do not extend this analysis to the `West' region as the field associated with this region is possibly an extension from the LMC rather than the SMC. We now explore this possibility.

\subsubsection{The LMC-West Field}

Figure\,\ref{fig:corrRMs} shows a nearly consistent, negative $\phi$-value in the region nearest the LMC. Leading to this region, \citet{Gaensler2005} show the bulk of the polarized sources in the nearest portion of the LMC also have negative RM-values\footnotemark after MW-foreground correction. Contrasting this, the majority of the RMs associated with the LMC have positive values. 
\footnotetext{Previous studies explicitly use the term rotation measure (RM), rather than Faraday depth, to express the magnitude of observed Faraday rotation.}

In the plane-of-the-sky magnetic field, both \citet{Wayte1990} and \citet{Mao2012} note that there are regions of the LMC where the magnetic field appears to align with the direction towards the SMC. 

If the MB-field is built from the magnetized material that originated in both Magellanic Clouds, the LMC contribution is likely associated with the tidal filaments ($l\,\simeq\,285.7, b\,\simeq\,-33.7$), first identified by \citet{Hayes1991}. \citet{Mao2012} estimate that the tidal filaments stemming from the LMC have a magnetic-field strength of $B_{T}\,=\,11\,\mu$G. They argue that the similar magnitude of off-source and on-source Faraday depths signifies that the line-of-sight magnetic field in this region is negligible. 

We measure an average $\overline{\phi}_{MB}\,=\,-21\,$rad\,m$^{-2}$ in the region of the filament discussed above. Although we have fewer off-source points, we find a scatter of the nearest 10 sources to be $\sigma\sim2\,$rad\,m$^{-2}$. For there to be no line-of-sight magnetic field in this region, the average electron density would have to be zero, which is unphysical. We estimate the $B_{\parallel}$ of the tidal filament using the same estimated pathlength as \citet{Mao2012}, $L_{H\textsc{ii}}\,=\,800\,$pc and Case 1 ($\S$\ref{sec:models}). The implied line-of-sight magnetic field in this region is $B_{\parallel}\,=\,1.2\,\mu$G, a value that is twice as strong as our initial estimates ($L_{H\textsc{ii}}\,=\,fL_5$) for the same region. 

This exercise demonstrates two things: firstly, it is possible that the Faraday depth values we see in the region nearest the LMC could be a consequence of a coherent magnetic field having been stripped from the LMC. Secondly, with different pathlength estimates, the magnetic field contribution from the LMC could be much higher than our initial estimates, implying a stronger total magnetic field strength in the MB.

\subsubsection{Does the pan-Magellanic Field Exist?}
\label{sec:pM_field}

We have discussed the implications of the known magnetic fields of the SMC and LMC as they pertain to our observed $\phi_{\rm{MB}}$ and $B_{\parallel, MB}$ in an attempt to justify the assumption that the observed MB magnetic field originated from both galaxies. We posit that the dominant magnetic field component is in the plane of the sky, a claim that is consistent with what has been previously argued in all discussions of the pan-Magellanic field. 

However, the previous discussions and implications of the SMC- and LMC magnetic fields assumed that the observed MB field is a combination of magnetic fields that have been drawn out of the LMC and SMC. Below, we briefly explore if the observed coherent field in the MB could have been formed in situ.

The $\alpha-\omega$ dynamo -- which is believed to be the mechanism responsible for the observation of coherent magnetic fields on the scales of galaxies -- requires too large a timescale to explain the existence of a coherent field in the young MB. By comparison, the cosmic-ray driven dynamo works on much shorter timescales. However, both of these dynamos require there to be differential rotation in the MB, which has not been observed. Therefore, these mechanisms cannot be responsible for the magnetic field in the MB. By contrast, the typical amplification time of the fluctuating dynamo is $10^6\,-\,10^7$\,yr, a timescale that is favourable given the age of the MB. However, this mechanism creates turbulent or incoherent magnetic fields and cannot be responsible for the observed ordered field in the MB.

While we do not address the origins of the magnetic fields in the LMC and SMC, the standard magnetic-field creation mechanisms cannot explain the existence of a coherent magnetic field in the young tidal remnant. We therefore conclude that the magnetic field in the MB is a consequence of field lines having been dragged out of the LMC and SMC with an overarching field geometry that has been determined by the orientation of the parent galaxies. This shared magnetic history links the two Magellanic Clouds and establishes the existence of the pan-Magellanic field.

All previous detections of tidal bridges with corresponding polarization have been detected through polarized continuum emission emanating from tidal regions (e.g. \citealt{Condon1993, Nikiel2013a}a \citealt{Nikiel2013b}b). Our detection of a coherent magnetic field in the MB was made using Faraday rotation observations through a non-continuum foreground, and is the first ever such detection for any tidal bridge. This may imply that magnetic fields are an early influence on the evolution of galaxy interactions. If magnetic fields do affect early galaxy interactions, the existence of magnetic fields in tidal remnants could explain the observation of coherent magnetic fields in tidal dwarfs (\citealt{Nikiel2013b}b) and would suggest the existence of magnetic fields in more diffuse tidal features, such as the Magellanic Stream and Leading Arm.

\section{Summary}
\label{sec:conclude}

We have presented Faraday rotation data for 167 extragalactic polarized sources and observe a coherent magnetic field towards the MB. Each source in our catalogue has well-determined polarization ($\mathcal{P}\,\geq\,8\sigma_{\mathcal{P}}$). Using a Monte Carlo Markov Chain approach to fitting observed complex polarization spectra, we were able to recover the polarization parameters of each source to high confidence.

We have demonstrated that the observed Faraday depths of sources `on' and `off' the MB are inherently different and have attributed this disparity to the existence of a large-scale, coherent magnetic field within the MB. We assumed a line-of-sight depth through the MB of $5$\,kpc and explored different distributions of ionized gas. The median line-of-sight magnetic field derived from these approximations are all consistent with $B_{\parallel}\,\simeq\,0.3\,\mu$G, where the uniform field is directed away from us. We stress that little is known about the distribution of ionized gas within the MB and the implied magnetic field is dependent upon this constraint. 

The MB is a tidal remnant that we argued has no known means for creating a coherent field on the scales observed. Therefore, we concluded that the magnetic fields of the LMC and SMC have been tidally stripped along with the neutral gas emanating from these galaxies to form the MB. The implied line-of-sight magnetic-field strength in the MB region nearest the SMC, which is where the majority of the gas of the MB is believed to have originated, is consistent with observed line-of-sight component of this galaxy. We have argued that the magnetic field associated with the LMC and its polarized filaments has also been pulled into the MB and are likely responsible for the observed Faraday-rotation in the region nearest these features.

This work represents the first observational confirmation of the pan-Magellanic field -- a coherent magnetic field spanning the entirety of the MB with a history and evolutionary fate that is tied to that of the Magellanic System.

\section{Acknowledgements}
J.\,F.\,K thanks Sui Ann Mao and Niels Oppermann for their useful discussions in the initial planning stages of this project. Furthermore, she graciously acknowledges the support of the Australia Telescope Compact Array staff for their hospitality during her stay at the facility. The Australia Telescope is funded by the Commonwealth of Australia for operation as a National Facility managed by CSIRO.
B.\,M.\,G. and C.\,R.\,P acknowledge the support of the Australian Research Council through grant FL100100114. The Dunlap Institute is funded through an endowment established by the David Dunlap family and the University of Toronto.
 N.\,M.\,M.-G. acknowledges the support of the Australian Research Council through Future Fellowship FT150100024.
 This research made use of Astropy, a community-developed core Python package for Astronomy \citep{Astropy}. This research has also made use of NASA's Astrophysics Data System. Many images for this paper were created using the Karma toolkit \citep{kvis}.


\bibliographystyle{mn2e}
\bibliography{allBibs}


\onecolumn
\appendix
\section{Table of Symbols}
\begin{longtable}{l l}
\caption{List of symbols used in this paper and their meaning.}\\
Symbol															&			Physical Quantity																																										\\[5pt]
\hline		
&\\													
$B_{\parallel}$											&			measured magnetic-field strength along the line-of-sight in units of $\mu$G 													\\[5pt]
$B_{\parallel}^{*}$									&			median magnetic-field strength along the line-of-sight in units of $\mu$G 														\\[5pt]
${B}_{c,T}$													&			total coherent magnetic-field strength, in units of $\mu$G																						\\[5pt]
$B_r$															&			random magnetic-field strength, in units of $\mu$G																									\\[5pt]
DM																	&			measured dispersion measure for a specific sightline in units of pc\,cm$^{-3}$												\\[5pt]
$\langle$EM$\rangle$							&			average emission measure for specified region, in units of pc\,cm$^{-6}$										\\[5pt]
EM																	&			measured emission measure along a specific sightline, pc\,cm$^{-6}$																\\[5pt]
$f$																	&			volume filling factor of gas such that the effective pathlength of gas with a characteristic density $n_0$ is $f\,\times\,n_0$																														\\[5pt]
$X$																	&			ionization fraction																														\\[5pt]
$I, Q, U, V$													&			observed Stokes parameters, with units of mJy																											\\[5pt]
$I_{\text{H}\alpha}$								&			intensity of H$\alpha$ emission, with units of rayleighs																							\\[5pt]
$L_5$															&			$5000$\,pc. The nominal line-of-sight depth of the MB.															\\[5pt]
$L_{\text{H\textsc{i}}}$							&			estimated line-of-sight depth of neutral hydrogen, in units of pc																			\\[5pt]
$L_{\text{H}^+}$										&			estimated line-of-sight depth of ionized material, in units of pc																			\\[5pt]
$\Delta\,L$													&			estimated standard deviation in line-of-sight depth of the MB, in units of pc					\\[5pt]
IQR																	&			inter-quartile range, defined as the range of values between the 25th and 75th percentiles						\\[5pt]
$l_0$																&			typical cell size along line-of-sight, in units of pc																											\\[5pt]
$\lambda^{2}$											&			the square of the observed wavelength, in units of m$^{2}$																					\\[5pt]
$\langle{N_{\text{H}\textsc{i}}}\rangle$			&			average H\textsc{i} column density, in units of cm$^{-2}$																												\\[5pt]
$\overline{n_{\text{H}\textsc{i}}}$	&			average neutral-gas density, calculated as $\langle{N_{\text{H}\textsc{i}}}\rangle/L_{\text{H}\textsc{i}}$ in units of cm$^{-3}$		\\[5pt]
$\overline{n_e}$										&			average free electron density, in units of cm$^{-3}$																												\\[5pt]
$\mathcal{P}$											&			polarized intensity in units of mJy/beam																															\\[5pt]
$p$																	&			observed polarized fraction																								\\[5pt]
$p_0$															& 		intrinsic polarized fraction																																\\[5pt]
$\phi_{\rm{corr}}$												&			Faraday depth for which the foreground, MW contribution has been subtracted, in units of rad\,m$^{-2}$		\\[5pt]
$\phi_{\rm{MB}}$												&			Faraday depth of the MB, in units of rad\,m$^{-2}$																		\\[5pt]
$\phi_{\rm{obs}}$												&			observed Faraday depth in units of rad\,m$^{-2}$																			 \\[5pt]
$d\phi$														&		 	error estimate in $\phi$ from fitting algorithm, $qu$-fitting	, in units of rad\,m$^{-2}$				\\[5pt]
$\overline{\phi}$										&			mean Faraday depth in units of rad\,m$^{-2}$																												\\[5pt]	
$\sigma^2_{\phi}$									&			variance of Faraday depths on scales smaller than the synthesised beam, in units of rad\,m$^{-2}$			\\[5pt]
$\sigma(\phi)$											&			standard deviation of an ensemble of Faraday depth values for a specified region, in units of rad\,m$^{-2}$							\\[5pt]
$\Psi$															& 		observed polarization angle, defined as $0.5\,arctan\frac{U}{Q}$																			\\[5pt]
$\Psi_0$														& 		intrinsic polarization angle at the source of emission																									\\[5pt]
$\mathcal{Q}_1$, $\mathcal{Q}_2$	&			first and third quartile, defined as the 25th and 75th population percentile value										\\[5pt]
RM																	&			the classical rotation measure, defined as $(\Psi_{0} - \Psi)/\lambda^2$, in units of rad\,m$^{-2}$		\\[5pt]
$\sigma_{\mathcal{P}}$						&			rms error in extracted polarized intensity, in units of mJy\,b$^{-1}$																			\\[5pt] 
$q, u$															&			fractional linear polarized Stokes $Q$ and $U$ parameters, units of per cent 												\\[5pt]		
$T_4$															&			assumed temperature of $10^4\,$K for the ionized medium										 												\\[5pt]												

\end{longtable}


\clearpage
\section{Table of derived Faraday depths}
\label{sec:apndxB}
\begin{longtable}{cc c | c c c c | c }
\caption{Table of the Faraday depth values for each polarized source used in our analysis. Sources proceeded by a '*' indicate targets that are considered to be off the Bridge, whereas those without an asterisk are considered to be on-Bridge sources. Columns 1 and 2 give the position of the source in galactic longitude and latitude. Columns 3-6 lists the best-fit values returned from the $q-u$ fitting routine, namely the intrinsic polarization fraction, intrinsic polarization angle, observed Faraday depth and Faraday dispersion. Each of the the uncertainties represents the 1$\sigma$ standard deviation in parameter space. Column 7 lists the corresponding Faraday depth of each source once the Faraday rotation due to the Milky Way foreground has been corrected for. The uncertainty calculation for this value is described in detail $\S$4.1.}\\
& $l$					&	$b$ 				&	 $p_0$	&		$\Psi_0$			&		$\phi_{raw}$		&		$\sigma_{\phi}$	&		$\phi_{\rm{corr}}$ \\[2mm]
& ($^{\circ}$)	&	($^{\circ}$)	&		(\%)	&		($^{\circ}$)		&		(rad\,m$^{-2}$)	&		(rad\,m$^{-2}$)		&		(rad\,m$^{-2}$)	 \\[1mm]
\hline\hline
\endhead
&	*	282.073	&	-42.586	&	0.025	$\pm$	0.001	&	10.5	$\pm$	2.7	&	24.9	$\pm$	1.5	&	0.9	$\pm$	0.8	&	-1.2	$\pm$	1.5	\\
&	*	283.003	&	-45.398	&	0.068	$\pm$	0.001	&	3.1	$\pm$	0.7	&	47.9	$\pm$	0.5	&	11.0	$\pm$	0.5	&	25.9	$\pm$	0.5	\\
&	*	283.340	&	-41.945	&	0.043	$\pm$	0.001	&	44.0	$\pm$	1.9	&	24.2	$\pm$	1.1	&	1.0	$\pm$	0.9	&	-2.0	$\pm$	1.1	\\
&		283.601	&	-33.891	&	0.078	$\pm$	0.004	&	0.9	$\pm$	2.7	&	47.8	$\pm$	1.5	&	10.0	$\pm$	1.3	&	11.4	$\pm$	1.6	\\
&		283.747	&	-34.130	&	0.097	$\pm$	0.005	&	94.9	$\pm$	2.1	&	40.6	$\pm$	1.2	&	7.5	$\pm$	1.5	&	4.6	$\pm$	1.2	\\
&		283.937	&	-32.823	&	0.045	$\pm$	0.001	&	3.8	$\pm$	0.6	&	62.7	$\pm$	0.5	&	9.0	$\pm$	0.4	&	25.0	$\pm$	0.6	\\
&		283.953	&	-33.160	&	0.060	$\pm$	0.002	&	16.5	$\pm$	1.9	&	25.9	$\pm$	1.1	&	1.1	$\pm$	1.0	&	-11.3	$\pm$	1.1	\\
&	*	284.041	&	-45.740	&	0.044	$\pm$	0.001	&	6.2	$\pm$	1.0	&	25.7	$\pm$	0.7	&	16.6	$\pm$	0.4	&	4.6	$\pm$	0.7	\\
&		284.078	&	-36.344	&	0.073	$\pm$	0.004	&	19.9	$\pm$	2.5	&	10.2	$\pm$	1.5	&	11.6	$\pm$	1.1	&	-22.9	$\pm$	1.5	\\
&		284.178	&	-35.180	&	0.072	$\pm$	0.002	&	14.7	$\pm$	1.6	&	3.8	$\pm$	0.9	&	1.3	$\pm$	1.1	&	-30.7	$\pm$	0.9	\\
&		284.193	&	-35.940	&	0.032	$\pm$	0.002	&	93.9	$\pm$	2.2	&	32.5	$\pm$	1.6	&	9.7	$\pm$	1.5	&	-1.0	$\pm$	1.6	\\
&		284.245	&	-35.661	&	0.162	$\pm$	0.015	&	118.6	$\pm$	4.0	&	24.6	$\pm$	2.3	&	13.3	$\pm$	1.6	&	-9.3	$\pm$	2.4	\\
&		284.904	&	-37.103	&	0.070	$\pm$	0.008	&	121.7	$\pm$	5.4	&	22.8	$\pm$	3.2	&	14.4	$\pm$	1.8	&	-8.9	$\pm$	3.2	\\
&		285.180	&	-33.102	&	0.053	$\pm$	0.004	&	43.2	$\pm$	2.9	&	21.4	$\pm$	1.9	&	14.5	$\pm$	1.2	&	-15.2	$\pm$	1.9	\\
&		285.244	&	-35.736	&	0.060	$\pm$	0.003	&	16.8	$\pm$	1.9	&	12.8	$\pm$	1.0	&	5.9	$\pm$	1.6	&	-20.4	$\pm$	1.1	\\
&		285.485	&	-31.527	&	0.334	$\pm$	0.009	&	44.5	$\pm$	1.2	&	37.0	$\pm$	0.7	&	7.5	$\pm$	0.7	&	-1.5	$\pm$	0.7	\\
&		285.532	&	-35.854	&	0.133	$\pm$	0.006	&	42.2	$\pm$	2.0	&	24.0	$\pm$	1.1	&	8.0	$\pm$	1.1	&	-8.9	$\pm$	1.1	\\
&		285.621	&	-37.178	&	0.132	$\pm$	0.007	&	122.4	$\pm$	2.5	&	-1.7	$\pm$	1.4	&	11.6	$\pm$	1.0	&	-32.9	$\pm$	1.4	\\
&	*	285.625	&	-39.347	&	0.064	$\pm$	0.001	&	143.9	$\pm$	0.6	&	26.8	$\pm$	0.3	&	13.9	$\pm$	0.2	&	-1.6	$\pm$	0.3	\\
&	*	285.970	&	-40.392	&	0.075	$\pm$	0.007	&	87.3	$\pm$	3.6	&	8.7	$\pm$	2.0	&	7.4	$\pm$	2.6	&	-18.2	$\pm$	2.0	\\
&		286.019	&	-37.718	&	0.049	$\pm$	0.002	&	134.3	$\pm$	1.9	&	16.3	$\pm$	1.4	&	15.6	$\pm$	0.8	&	-14.0	$\pm$	1.4	\\
&		286.022	&	-37.647	&	0.248	$\pm$	0.011	&	12.3	$\pm$	2.9	&	27.5	$\pm$	1.4	&	2.5	$\pm$	1.9	&	-2.9	$\pm$	1.4	\\
&	*	286.253	&	-45.332	&	0.400	$\pm$	0.035	&	177.6	$\pm$	7.5	&	40.6	$\pm$	3.4	&	3.6	$\pm$	2.5	&	20.2	$\pm$	3.4	\\
&		286.352	&	-32.410	&	0.158	$\pm$	0.018	&	147.1	$\pm$	4.2	&	10.6	$\pm$	2.0	&	8.8	$\pm$	2.4	&	-26.3	$\pm$	2.1	\\
&		286.582	&	-34.181	&	0.047	$\pm$	0.001	&	154.4	$\pm$	0.7	&	6.6	$\pm$	0.4	&	1.3	$\pm$	1.0	&	-27.9	$\pm$	0.5	\\
&	*	286.660	&	-41.685	&	0.101	$\pm$	0.002	&	118.0	$\pm$	1.1	&	13.3	$\pm$	0.6	&	1.8	$\pm$	1.3	&	-11.6	$\pm$	0.6	\\
&		286.672	&	-31.294	&	0.291	$\pm$	0.013	&	124.5	$\pm$	2.4	&	42.8	$\pm$	2.3	&	30.6	$\pm$	0.9	&	4.6	$\pm$	2.3	\\
&	*	286.782	&	-45.866	&	0.076	$\pm$	0.001	&	109.4	$\pm$	0.8	&	14.4	$\pm$	0.4	&	3.5	$\pm$	1.1	&	-5.1	$\pm$	0.4	\\
&		286.858	&	-33.944	&	0.082	$\pm$	0.006	&	150.3	$\pm$	3.1	&	26.9	$\pm$	1.6	&	6.9	$\pm$	2.1	&	-7.8	$\pm$	1.6	\\
&		286.919	&	-34.667	&	0.324	$\pm$	0.024	&	71.1	$\pm$	3.2	&	0.9	$\pm$	1.7	&	11.5	$\pm$	1.3	&	-32.9	$\pm$	1.8	\\
&	*	287.005	&	-45.668	&	0.114	$\pm$	0.001	&	89.0	$\pm$	0.4	&	28.2	$\pm$	0.2	&	0.5	$\pm$	0.4	&	8.6	$\pm$	0.2	\\
&		287.005	&	-32.386	&	0.070	$\pm$	0.002	&	175.3	$\pm$	2.1	&	13.3	$\pm$	1.1	&	1.3	$\pm$	1.1	&	-23.3	$\pm$	1.1	\\
&	*	287.015	&	-45.653	&	0.117	$\pm$	0.001	&	88.9	$\pm$	0.4	&	27.6	$\pm$	0.2	&	3.8	$\pm$	0.5	&	7.9	$\pm$	0.2	\\
&	*	287.195	&	-41.776	&	0.070	$\pm$	0.006	&	73.3	$\pm$	3.8	&	9.0	$\pm$	2.3	&	10.9	$\pm$	2.0	&	-15.5	$\pm$	2.3	\\
&		287.249	&	-33.344	&	0.094	$\pm$	0.003	&	130.9	$\pm$	1.3	&	0.2	$\pm$	0.7	&	11.5	$\pm$	0.5	&	-35.0	$\pm$	0.8	\\
&		287.439	&	-37.078	&	0.042	$\pm$	0.001	&	160.9	$\pm$	1.4	&	14.4	$\pm$	0.8	&	9.6	$\pm$	0.7	&	-16.0	$\pm$	0.8	\\
&	*	287.529	&	-45.345	&	0.058	$\pm$	0.001	&	150.2	$\pm$	1.0	&	16.2	$\pm$	0.7	&	13.4	$\pm$	0.4	&	-3.5	$\pm$	0.7	\\
&	*	287.545	&	-43.624	&	0.087	$\pm$	0.002	&	3.8	$\pm$	1.0	&	7.9	$\pm$	0.6	&	10.3	$\pm$	0.5	&	-14.0	$\pm$	0.6	\\
&		287.675	&	-35.482	&	0.118	$\pm$	0.007	&	159.4	$\pm$	2.6	&	31.1	$\pm$	1.6	&	12.4	$\pm$	1.1	&	-1.3	$\pm$	1.6	\\
&	*	288.067	&	-45.870	&	0.111	$\pm$	0.002	&	146.6	$\pm$	1.2	&	9.0	$\pm$	0.7	&	0.9	$\pm$	0.8	&	-9.8	$\pm$	0.7	\\
&		288.421	&	-37.393	&	0.067	$\pm$	0.003	&	164.6	$\pm$	2.3	&	-30.1	$\pm$	1.3	&	4.7	$\pm$	2.2	&	-59.6	$\pm$	1.3	\\
&		288.484	&	-33.311	&	0.139	$\pm$	0.008	&	111.5	$\pm$	4.0	&	18.8	$\pm$	2.1	&	2.3	$\pm$	1.9	&	-15.9	$\pm$	2.2	\\
&		288.589	&	-39.501	&	0.017	$\pm$	0.001	&	81.8	$\pm$	3.1	&	-0.2	$\pm$	1.8	&	2.9	$\pm$	2.2	&	-26.9	$\pm$	1.8	\\
&	*	288.627	&	-41.413	&	0.084	$\pm$	0.004	&	133.5	$\pm$	2.1	&	17.5	$\pm$	2.0	&	28.2	$\pm$	1.0	&	-6.8	$\pm$	2.0	\\
&	*	288.715	&	-40.820	&	0.964	$\pm$	0.027	&	10.7	$\pm$	4.8	&	22.5	$\pm$	4.3	&	32.5	$\pm$	1.1	&	-2.5	$\pm$	4.3	\\
&		289.090	&	-39.342	&	0.127	$\pm$	0.006	&	49.2	$\pm$	2.1	&	13.8	$\pm$	1.2	&	8.2	$\pm$	1.2	&	-12.9	$\pm$	1.2	\\
&		289.105	&	-33.833	&	0.048	$\pm$	0.004	&	30.0	$\pm$	3.7	&	24.6	$\pm$	2.6	&	15.7	$\pm$	1.9	&	-9.1	$\pm$	2.6	\\
&	*	289.145	&	-44.971	&	0.080	$\pm$	0.007	&	174.8	$\pm$	4.8	&	48.2	$\pm$	4.1	&	18.4	$\pm$	2.1	&	28.8	$\pm$	4.1	\\
&		289.161	&	-32.618	&	0.112	$\pm$	0.007	&	28.2	$\pm$	2.5	&	37.1	$\pm$	1.3	&	6.9	$\pm$	1.8	&	1.8	$\pm$	1.3	\\
&		289.167	&	-32.625	&	0.111	$\pm$	0.006	&	35.3	$\pm$	2.4	&	34.9	$\pm$	1.4	&	10.2	$\pm$	1.3	&	-0.3	$\pm$	1.4	\\
&	*	289.253	&	-45.636	&	0.168	$\pm$	0.018	&	83.4	$\pm$	4.4	&	9.0	$\pm$	2.7	&	11.1	$\pm$	2.1	&	-9.5	$\pm$	2.7	\\
&		289.395	&	-32.792	&	0.132	$\pm$	0.006	&	120.5	$\pm$	1.8	&	32.1	$\pm$	1.0	&	5.8	$\pm$	1.6	&	-2.8	$\pm$	1.0	\\
&		289.478	&	-39.488	&	0.037	$\pm$	0.002	&	170.7	$\pm$	2.9	&	4.4	$\pm$	1.5	&	2.9	$\pm$	2.2	&	-21.9	$\pm$	1.5	\\
&	*	290.012	&	-41.763	&	0.051	$\pm$	0.001	&	169.3	$\pm$	0.7	&	7.9	$\pm$	0.4	&	10.5	$\pm$	0.3	&	-15.2	$\pm$	0.4	\\
&		290.038	&	-39.129	&	0.044	$\pm$	0.004	&	57.1	$\pm$	3.4	&	4.9	$\pm$	2.6	&	19.9	$\pm$	1.5	&	-21.5	$\pm$	2.6	\\
&		290.434	&	-36.118	&	0.165	$\pm$	0.001	&	43.0	$\pm$	0.5	&	23.5	$\pm$	0.3	&	0.3	$\pm$	0.3	&	-6.6	$\pm$	0.3	\\
&		290.710	&	-38.878	&	0.170	$\pm$	0.017	&	111.1	$\pm$	4.0	&	14.8	$\pm$	2.3	&	12.8	$\pm$	1.7	&	-11.6	$\pm$	2.3	\\
&		290.754	&	-38.330	&	0.038	$\pm$	0.001	&	21.3	$\pm$	1.5	&	21.6	$\pm$	0.8	&	2.4	$\pm$	1.7	&	-5.5	$\pm$	0.8	\\
&		290.754	&	-35.106	&	0.074	$\pm$	0.001	&	107.8	$\pm$	1.1	&	4.2	$\pm$	0.6	&	0.6	$\pm$	0.6	&	-27.0	$\pm$	0.6	\\
&		290.852	&	-32.259	&	0.051	$\pm$	0.002	&	172.8	$\pm$	1.6	&	40.2	$\pm$	0.8	&	6.7	$\pm$	1.1	&	5.4	$\pm$	0.8	\\
&	*	290.958	&	-45.418	&	0.028	$\pm$	0.003	&	49.3	$\pm$	4.6	&	13.1	$\pm$	4.4	&	22.7	$\pm$	1.9	&	-4.8	$\pm$	4.4	\\
&		290.961	&	-40.949	&	0.130	$\pm$	0.006	&	158.0	$\pm$	2.1	&	3.7	$\pm$	1.1	&	5.7	$\pm$	1.8	&	-19.9	$\pm$	1.1	\\
&		290.986	&	-36.119	&	0.058	$\pm$	0.002	&	35.5	$\pm$	1.3	&	2.0	$\pm$	0.8	&	11.0	$\pm$	0.6	&	-27.8	$\pm$	0.8	\\
&		291.157	&	-38.395	&	0.127	$\pm$	0.001	&	3.6	$\pm$	0.4	&	11.4	$\pm$	0.2	&	1.6	$\pm$	0.9	&	-15.4	$\pm$	0.3	\\
&		291.468	&	-44.002	&	0.536	$\pm$	0.026	&	90.4	$\pm$	3.2	&	6.2	$\pm$	1.9	&	2.3	$\pm$	1.9	&	-13.2	$\pm$	1.9	\\
&		291.492	&	-41.596	&	0.033	$\pm$	0.001	&	179.4	$\pm$	1.2	&	11.3	$\pm$	0.6	&	5.6	$\pm$	0.9	&	-11.3	$\pm$	0.6	\\
&		291.508	&	-40.460	&	0.199	$\pm$	0.005	&	54.3	$\pm$	1.3	&	0.3	$\pm$	0.8	&	11.8	$\pm$	0.5	&	-23.7	$\pm$	0.8	\\
&		291.652	&	-34.097	&	0.079	$\pm$	0.006	&	175.1	$\pm$	5.0	&	11.5	$\pm$	5.7	&	39.8	$\pm$	2.0	&	-20.6	$\pm$	5.7	\\
&		291.674	&	-34.518	&	0.044	$\pm$	0.001	&	48.1	$\pm$	0.6	&	23.2	$\pm$	0.4	&	0.5	$\pm$	0.5	&	-8.4	$\pm$	0.4	\\
&		291.778	&	-40.785	&	0.059	$\pm$	0.007	&	37.7	$\pm$	4.6	&	6.0	$\pm$	3.6	&	21.1	$\pm$	2.2	&	-17.4	$\pm$	3.6	\\
&		291.865	&	-37.269	&	0.134	$\pm$	0.010	&	91.7	$\pm$	2.9	&	37.7	$\pm$	2.6	&	26.7	$\pm$	1.5	&	9.9	$\pm$	2.6	\\
&		291.865	&	-37.269	&	0.114	$\pm$	0.008	&	95.2	$\pm$	2.8	&	33.8	$\pm$	2.3	&	27.6	$\pm$	1.3	&	5.9	$\pm$	2.3	\\
&		291.881	&	-35.986	&	0.075	$\pm$	0.004	&	58.2	$\pm$	2.3	&	12.9	$\pm$	1.2	&	6.7	$\pm$	1.6	&	-16.6	$\pm$	1.2	\\
&	*	291.954	&	-31.336	&	0.161	$\pm$	0.006	&	87.6	$\pm$	2.0	&	38.1	$\pm$	1.4	&	19.3	$\pm$	0.7	&	2.7	$\pm$	1.4	\\
&		292.041	&	-43.665	&	0.193	$\pm$	0.010	&	50.6	$\pm$	4.5	&	9.3	$\pm$	2.4	&	1.4	$\pm$	1.2	&	-10.3	$\pm$	2.4	\\
&		292.082	&	-42.348	&	0.186	$\pm$	0.012	&	26.8	$\pm$	2.7	&	12.2	$\pm$	1.4	&	7.7	$\pm$	1.8	&	-9.0	$\pm$	1.4	\\
&		292.397	&	-42.859	&	0.128	$\pm$	0.007	&	103.2	$\pm$	3.5	&	-1.2	$\pm$	1.8	&	3.0	$\pm$	2.2	&	-21.7	$\pm$	1.8	\\
&		292.473	&	-37.351	&	0.018	$\pm$	0.000	&	129.6	$\pm$	1.1	&	23.6	$\pm$	0.6	&	3.9	$\pm$	1.5	&	-3.9	$\pm$	0.6	\\
&		292.550	&	-41.942	&	0.545	$\pm$	0.026	&	65.5	$\pm$	2.1	&	9.1	$\pm$	1.3	&	13.2	$\pm$	0.8	&	-12.5	$\pm$	1.3	\\
&		292.652	&	-44.236	&	0.293	$\pm$	0.003	&	19.9	$\pm$	0.4	&	18.2	$\pm$	0.3	&	15.8	$\pm$	0.2	&	-0.4	$\pm$	0.4	\\
&		292.741	&	-43.379	&	0.235	$\pm$	0.019	&	87.9	$\pm$	3.5	&	15.9	$\pm$	1.8	&	7.3	$\pm$	2.3	&	-3.7	$\pm$	1.8	\\
&	*	292.833	&	-31.083	&	0.081	$\pm$	0.003	&	35.4	$\pm$	2.4	&	40.7	$\pm$	1.2	&	2.1	$\pm$	1.7	&	5.4	$\pm$	1.2	\\
&		292.868	&	-41.315	&	0.320	$\pm$	0.034	&	175.4	$\pm$	4.6	&	4.4	$\pm$	2.8	&	7.9	$\pm$	3.5	&	-17.8	$\pm$	2.8	\\
&		293.055	&	-39.953	&	0.146	$\pm$	0.012	&	64.2	$\pm$	3.7	&	-0.8	$\pm$	2.0	&	10.2	$\pm$	1.6	&	-24.7	$\pm$	2.0	\\
&		293.088	&	-43.924	&	0.363	$\pm$	0.033	&	78.7	$\pm$	5.0	&	27.7	$\pm$	3.9	&	8.3	$\pm$	4.9	&	9.0	$\pm$	3.9	\\
&		293.148	&	-41.268	&	0.089	$\pm$	0.002	&	13.7	$\pm$	1.5	&	-20.0	$\pm$	1.1	&	19.4	$\pm$	0.5	&	-42.1	$\pm$	1.1	\\
&		293.471	&	-42.911	&	0.178	$\pm$	0.007	&	122.4	$\pm$	1.7	&	10.0	$\pm$	1.0	&	11.2	$\pm$	0.7	&	-9.8	$\pm$	1.0	\\
&		293.509	&	-43.572	&	0.317	$\pm$	0.014	&	80.4	$\pm$	1.9	&	20.6	$\pm$	1.1	&	7.9	$\pm$	1.2	&	1.6	$\pm$	1.1	\\
&		293.635	&	-44.131	&	0.229	$\pm$	0.013	&	25.0	$\pm$	2.4	&	31.3	$\pm$	1.5	&	16.1	$\pm$	0.9	&	13.1	$\pm$	1.5	\\
&		293.737	&	-42.296	&	0.284	$\pm$	0.018	&	130.6	$\pm$	2.8	&	10.9	$\pm$	1.6	&	6.9	$\pm$	2.2	&	-9.6	$\pm$	1.6	\\
&		293.807	&	-41.469	&	0.129	$\pm$	0.012	&	61.1	$\pm$	3.9	&	13.5	$\pm$	2.3	&	11.8	$\pm$	1.7	&	-8.0	$\pm$	2.3	\\
&		293.819	&	-41.466	&	0.154	$\pm$	0.012	&	71.9	$\pm$	3.3	&	6.4	$\pm$	2.0	&	14.2	$\pm$	1.4	&	-15.1	$\pm$	2.0	\\
&	*	293.851	&	-31.371	&	0.035	$\pm$	0.000	&	98.3	$\pm$	0.3	&	43.8	$\pm$	0.2	&	0.1	$\pm$	0.1	&	9.4	$\pm$	0.2	\\
&		293.907	&	-39.279	&	0.385	$\pm$	0.021	&	53.0	$\pm$	4.5	&	17.9	$\pm$	2.1	&	2.4	$\pm$	2.0	&	-6.3	$\pm$	2.1	\\
&		294.006	&	-44.374	&	0.200	$\pm$	0.005	&	4.1	$\pm$	1.2	&	8.6	$\pm$	0.6	&	3.9	$\pm$	1.5	&	-9.1	$\pm$	0.7	\\
&		294.192	&	-43.969	&	0.234	$\pm$	0.019	&	156.9	$\pm$	3.5	&	17.8	$\pm$	1.9	&	6.4	$\pm$	2.8	&	-0.3	$\pm$	1.9	\\
&		294.271	&	-44.918	&	0.095	$\pm$	0.001	&	158.1	$\pm$	0.5	&	-8.4	$\pm$	0.3	&	3.1	$\pm$	0.7	&	-25.2	$\pm$	0.4	\\
&		294.385	&	-44.286	&	0.084	$\pm$	0.003	&	93.6	$\pm$	1.7	&	22.4	$\pm$	1.0	&	9.5	$\pm$	0.9	&	4.7	$\pm$	1.0	\\
&	*	294.480	&	-31.056	&	0.132	$\pm$	0.002	&	165.5	$\pm$	1.0	&	42.5	$\pm$	0.5	&	0.9	$\pm$	0.8	&	8.0	$\pm$	0.5	\\
&		294.525	&	-40.882	&	0.081	$\pm$	0.004	&	99.8	$\pm$	1.9	&	2.4	$\pm$	1.1	&	8.3	$\pm$	1.2	&	-19.5	$\pm$	1.1	\\
&		294.530	&	-42.244	&	0.347	$\pm$	0.024	&	12.2	$\pm$	3.6	&	18.4	$\pm$	1.8	&	4.3	$\pm$	2.7	&	-1.7	$\pm$	1.8	\\
&		294.535	&	-40.875	&	0.063	$\pm$	0.002	&	95.2	$\pm$	1.5	&	4.9	$\pm$	0.9	&	5.7	$\pm$	1.4	&	-17.0	$\pm$	1.0	\\
&		294.536	&	-43.701	&	0.454	$\pm$	0.019	&	53.2	$\pm$	1.8	&	21.9	$\pm$	0.9	&	7.9	$\pm$	1.0	&	3.6	$\pm$	0.9	\\
&		294.548	&	-40.966	&	0.107	$\pm$	0.007	&	21.2	$\pm$	3.1	&	4.7	$\pm$	1.6	&	5.0	$\pm$	2.4	&	-17.1	$\pm$	1.6	\\
&		294.593	&	-40.497	&	0.231	$\pm$	0.007	&	136.5	$\pm$	1.8	&	12.4	$\pm$	1.0	&	2.9	$\pm$	1.9	&	-10.0	$\pm$	1.0	\\
&		294.608	&	-42.908	&	0.060	$\pm$	0.007	&	161.1	$\pm$	4.2	&	12.2	$\pm$	2.5	&	8.7	$\pm$	3.4	&	-7.1	$\pm$	2.5	\\
&		294.713	&	-43.612	&	0.614	$\pm$	0.039	&	79.3	$\pm$	2.7	&	18.6	$\pm$	1.6	&	14.7	$\pm$	0.9	&	0.2	$\pm$	1.6	\\
&		294.884	&	-44.122	&	0.052	$\pm$	0.001	&	73.1	$\pm$	1.5	&	26.3	$\pm$	0.8	&	2.8	$\pm$	1.8	&	8.8	$\pm$	0.9	\\
&		294.912	&	-40.650	&	0.027	$\pm$	0.002	&	21.5	$\pm$	5.5	&	18.3	$\pm$	2.9	&	4.0	$\pm$	2.9	&	-3.7	$\pm$	2.9	\\
&		294.930	&	-42.244	&	0.065	$\pm$	0.002	&	157.6	$\pm$	2.6	&	21.7	$\pm$	1.4	&	1.8	$\pm$	1.5	&	1.8	$\pm$	1.4	\\
&		294.995	&	-41.978	&	0.051	$\pm$	0.002	&	49.8	$\pm$	1.7	&	25.5	$\pm$	1.1	&	9.5	$\pm$	1.1	&	5.3	$\pm$	1.1	\\
&		295.126	&	-41.492	&	0.057	$\pm$	0.003	&	109.3	$\pm$	2.0	&	35.5	$\pm$	1.1	&	5.1	$\pm$	1.9	&	14.7	$\pm$	1.1	\\
&		295.202	&	-41.984	&	0.167	$\pm$	0.014	&	86.4	$\pm$	3.5	&	28.6	$\pm$	2.0	&	9.8	$\pm$	1.8	&	8.4	$\pm$	2.0	\\
&		295.226	&	-42.823	&	0.294	$\pm$	0.015	&	92.6	$\pm$	2.2	&	-9.5	$\pm$	1.2	&	11.6	$\pm$	1.0	&	-28.6	$\pm$	1.3	\\
&		295.361	&	-40.387	&	0.121	$\pm$	0.007	&	54.4	$\pm$	4.1	&	25.0	$\pm$	2.2	&	2.6	$\pm$	2.2	&	2.9	$\pm$	2.2	\\
&		295.367	&	-40.785	&	0.322	$\pm$	0.041	&	147.4	$\pm$	10.5	&	22.9	$\pm$	5.1	&	2.9	$\pm$	2.6	&	1.3	$\pm$	5.1	\\
&		295.507	&	-42.082	&	0.162	$\pm$	0.009	&	86.8	$\pm$	2.2	&	27.7	$\pm$	1.2	&	7.2	$\pm$	1.7	&	7.8	$\pm$	1.2	\\
&		295.524	&	-40.997	&	0.062	$\pm$	0.006	&	85.3	$\pm$	6.4	&	11.5	$\pm$	3.1	&	3.7	$\pm$	2.9	&	-9.8	$\pm$	3.1	\\
&		295.675	&	-44.323	&	0.077	$\pm$	0.007	&	173.7	$\pm$	5.5	&	10.3	$\pm$	2.8	&	5.3	$\pm$	3.0	&	-6.6	$\pm$	2.8	\\
&	*	295.688	&	-34.852	&	0.130	$\pm$	0.003	&	9.9	$\pm$	1.4	&	30.8	$\pm$	0.9	&	5.7	$\pm$	1.1	&	1.8	$\pm$	0.9	\\
&		295.733	&	-42.013	&	0.142	$\pm$	0.010	&	105.0	$\pm$	3.4	&	34.9	$\pm$	2.0	&	5.8	$\pm$	2.7	&	15.1	$\pm$	2.0	\\
&		295.814	&	-43.417	&	0.274	$\pm$	0.030	&	174.5	$\pm$	5.0	&	21.8	$\pm$	2.8	&	8.3	$\pm$	3.2	&	3.8	$\pm$	2.8	\\
&		295.881	&	-43.599	&	0.022	$\pm$	0.001	&	72.5	$\pm$	2.7	&	42.0	$\pm$	1.8	&	15.8	$\pm$	0.9	&	24.3	$\pm$	1.8	\\
&		295.893	&	-42.505	&	0.205	$\pm$	0.011	&	48.4	$\pm$	2.6	&	19.3	$\pm$	1.6	&	15.5	$\pm$	0.9	&	0.2	$\pm$	1.6	\\
&		295.925	&	-45.474	&	0.473	$\pm$	0.037	&	85.4	$\pm$	3.6	&	-5.1	$\pm$	2.3	&	17.5	$\pm$	1.1	&	-20.4	$\pm$	2.3	\\
&		295.925	&	-43.169	&	0.209	$\pm$	0.014	&	73.3	$\pm$	3.2	&	18.8	$\pm$	1.9	&	9.9	$\pm$	1.6	&	0.5	$\pm$	1.9	\\
&		295.956	&	-43.659	&	0.044	$\pm$	0.001	&	37.3	$\pm$	1.4	&	23.1	$\pm$	0.8	&	6.6	$\pm$	1.1	&	5.4	$\pm$	0.9	\\
&		295.963	&	-43.664	&	0.054	$\pm$	0.002	&	39.0	$\pm$	1.4	&	21.7	$\pm$	1.0	&	12.1	$\pm$	0.7	&	4.1	$\pm$	1.0	\\
&		295.986	&	-42.955	&	0.325	$\pm$	0.054	&	106.3	$\pm$	7.2	&	6.3	$\pm$	3.8	&	9.5	$\pm$	3.9	&	-12.2	$\pm$	3.8	\\
&		296.022	&	-42.163	&	0.070	$\pm$	0.004	&	50.4	$\pm$	3.2	&	2.5	$\pm$	3.4	&	31.1	$\pm$	1.4	&	-17.0	$\pm$	3.4	\\
&		296.491	&	-40.813	&	0.226	$\pm$	0.011	&	157.0	$\pm$	3.4	&	4.5	$\pm$	1.9	&	1.6	$\pm$	1.4	&	-16.4	$\pm$	1.9	\\
&		296.659	&	-45.653	&	0.101	$\pm$	0.004	&	70.9	$\pm$	1.6	&	-13.1	$\pm$	1.0	&	10.6	$\pm$	0.8	&	-27.8	$\pm$	1.0	\\
&		296.704	&	-43.455	&	0.057	$\pm$	0.006	&	168.7	$\pm$	3.9	&	6.0	$\pm$	2.4	&	9.2	$\pm$	2.7	&	-11.5	$\pm$	2.4	\\
&		296.719	&	-40.842	&	0.194	$\pm$	0.004	&	55.0	$\pm$	0.8	&	4.7	$\pm$	0.5	&	6.4	$\pm$	0.6	&	-16.1	$\pm$	0.5	\\
&		296.882	&	-40.719	&	0.237	$\pm$	0.028	&	52.9	$\pm$	4.7	&	-7.1	$\pm$	2.3	&	8.5	$\pm$	2.8	&	-28.0	$\pm$	2.3	\\
&	*	296.933	&	-33.860	&	0.137	$\pm$	0.011	&	6.3	$\pm$	4.0	&	27.3	$\pm$	3.2	&	20.9	$\pm$	1.4	&	-2.4	$\pm$	3.2	\\
&		296.997	&	-40.395	&	0.063	$\pm$	0.002	&	68.3	$\pm$	1.4	&	1.3	$\pm$	0.7	&	2.6	$\pm$	1.6	&	-19.9	$\pm$	0.7	\\
&		297.068	&	-41.714	&	0.043	$\pm$	0.002	&	125.5	$\pm$	2.3	&	16.8	$\pm$	1.3	&	11.3	$\pm$	0.9	&	-2.8	$\pm$	1.3	\\
&		297.070	&	-41.711	&	0.042	$\pm$	0.002	&	114.8	$\pm$	2.7	&	20.2	$\pm$	1.5	&	10.8	$\pm$	1.0	&	0.7	$\pm$	1.5	\\
&		297.070	&	-41.257	&	0.063	$\pm$	0.003	&	145.0	$\pm$	3.4	&	18.4	$\pm$	1.8	&	1.7	$\pm$	1.5	&	-1.7	$\pm$	1.8	\\
&		297.257	&	-41.186	&	0.300	$\pm$	0.013	&	132.7	$\pm$	2.8	&	17.6	$\pm$	1.8	&	2.8	$\pm$	2.2	&	-2.5	$\pm$	1.8	\\
&		297.277	&	-42.705	&	0.628	$\pm$	0.056	&	9.5	$\pm$	6.1	&	11.9	$\pm$	3.7	&	4.6	$\pm$	3.9	&	-6.3	$\pm$	3.7	\\
&		297.344	&	-43.704	&	0.233	$\pm$	0.019	&	67.3	$\pm$	3.4	&	-18.4	$\pm$	2.1	&	8.6	$\pm$	2.2	&	-35.2	$\pm$	2.1	\\
&		297.476	&	-41.179	&	0.766	$\pm$	0.128	&	150.1	$\pm$	31.9	&	-21.9	$\pm$	12.1	&	5.1	$\pm$	3.4	&	-41.9	$\pm$	12.1	\\
&		297.624	&	-44.093	&	0.170	$\pm$	0.017	&	76.8	$\pm$	3.9	&	-11.0	$\pm$	2.1	&	8.3	$\pm$	2.6	&	-27.2	$\pm$	2.1	\\
&		297.758	&	-44.158	&	0.097	$\pm$	0.007	&	33.4	$\pm$	3.2	&	-14.2	$\pm$	2.1	&	15.5	$\pm$	1.4	&	-30.3	$\pm$	2.1	\\
&	*	298.169	&	-35.710	&	0.127	$\pm$	0.003	&	114.9	$\pm$	1.1	&	50.7	$\pm$	0.6	&	4.2	$\pm$	1.3	&	24.0	$\pm$	0.6	\\
&		298.209	&	-42.078	&	0.177	$\pm$	0.014	&	15.6	$\pm$	4.4	&	-15.7	$\pm$	2.4	&	3.7	$\pm$	2.7	&	-34.2	$\pm$	2.4	\\
&	*	298.247	&	-33.110	&	0.125	$\pm$	0.010	&	25.8	$\pm$	3.1	&	38.3	$\pm$	1.8	&	7.1	$\pm$	2.6	&	8.4	$\pm$	1.8	\\
&		298.251	&	-42.116	&	0.433	$\pm$	0.026	&	11.8	$\pm$	2.7	&	-46.6	$\pm$	1.6	&	7.7	$\pm$	1.9	&	-65.0	$\pm$	1.6	\\
&		298.381	&	-43.653	&	0.050	$\pm$	0.003	&	144.7	$\pm$	3.3	&	7.3	$\pm$	1.8	&	3.1	$\pm$	2.3	&	-9.1	$\pm$	1.8	\\
&	*	298.994	&	-36.643	&	0.031	$\pm$	0.002	&	135.7	$\pm$	2.8	&	18.3	$\pm$	2.0	&	15.9	$\pm$	1.1	&	-6.8	$\pm$	2.0	\\
&	*	299.493	&	-30.568	&	0.021	$\pm$	0.001	&	2.7	$\pm$	2.0	&	24.7	$\pm$	1.2	&	11.6	$\pm$	0.8	&	-7.9	$\pm$	1.2	\\
&	*	299.727	&	-33.232	&	0.019	$\pm$	0.001	&	178.4	$\pm$	2.5	&	36.5	$\pm$	1.5	&	13.4	$\pm$	0.9	&	7.5	$\pm$	1.5	\\
&		299.815	&	-41.686	&	0.190	$\pm$	0.003	&	23.6	$\pm$	1.4	&	38.4	$\pm$	1.0	&	0.9	$\pm$	0.8	&	20.2	$\pm$	1.0	\\
&	*	300.101	&	-32.957	&	0.107	$\pm$	0.002	&	42.9	$\pm$	1.5	&	19.3	$\pm$	0.8	&	1.3	$\pm$	1.1	&	-9.9	$\pm$	0.8	\\
&	*	300.176	&	-37.791	&	0.083	$\pm$	0.004	&	156.5	$\pm$	3.2	&	25.4	$\pm$	1.7	&	1.8	$\pm$	1.5	&	2.4	$\pm$	1.7	\\
&		300.260	&	-41.713	&	0.022	$\pm$	0.000	&	32.5	$\pm$	1.2	&	21.2	$\pm$	0.6	&	1.6	$\pm$	1.2	&	3.3	$\pm$	0.7	\\
&	*	300.546	&	-34.430	&	0.210	$\pm$	0.004	&	35.0	$\pm$	1.6	&	19.8	$\pm$	0.9	&	0.8	$\pm$	0.7	&	-7.3	$\pm$	0.9	\\
&	*	300.682	&	-38.509	&	0.098	$\pm$	0.004	&	124.7	$\pm$	1.6	&	31.4	$\pm$	1.0	&	11.6	$\pm$	0.7	&	9.6	$\pm$	1.0	\\
&	*	300.712	&	-31.222	&	0.023	$\pm$	0.001	&	26.9	$\pm$	1.6	&	12.4	$\pm$	0.9	&	5.8	$\pm$	1.3	&	-18.7	$\pm$	0.9	\\
&	*	300.977	&	-35.998	&	0.096	$\pm$	0.009	&	60.8	$\pm$	4.0	&	33.9	$\pm$	3.0	&	20.0	$\pm$	1.6	&	9.1	$\pm$	2.9	\\
&	*	301.077	&	-37.858	&	0.378	$\pm$	0.032	&	144.6	$\pm$	3.6	&	28.0	$\pm$	1.7	&	7.7	$\pm$	1.8	&	5.5	$\pm$	1.7	\\
&	*	301.241	&	-35.873	&	0.042	$\pm$	0.002	&	157.0	$\pm$	2.3	&	8.7	$\pm$	1.2	&	11.3	$\pm$	0.9	&	-16.2	$\pm$	1.2	\\
&	*	301.463	&	-32.596	&	0.089	$\pm$	0.003	&	37.9	$\pm$	1.3	&	28.7	$\pm$	0.8	&	10.9	$\pm$	0.7	&	-0.2	$\pm$	0.8	\\
&	*	302.602	&	-38.463	&	0.059	$\pm$	0.002	&	177.2	$\pm$	1.8	&	10.8	$\pm$	1.0	&	7.1	$\pm$	1.2	&	-10.1	$\pm$	1.0	\\
&	*	304.115	&	-35.992	&	0.041	$\pm$	0.003	&	138.4	$\pm$	4.0	&	48.2	$\pm$	2.8	&	15.4	$\pm$	1.5	&	25.0	$\pm$	2.8	\\
\end{longtable}

\end{document}